\documentclass[10pt,letterpaper,journal,twocolumn]{IEEEtran}
\usepackage{hyperref}
\usepackage{enumerate}
\usepackage{calc}
\usepackage{ifthen}

\usepackage{verbatim}
\usepackage[usenames,dvipsnames]{pstricks}
\usepackage{pst-node}   
\usepackage{epsfig}
 
\usepackage{amsmath}
\usepackage{latexsym}
\usepackage{ifthen}

\usepackage{amsfonts}
\usepackage{amsmath}
\usepackage{amssymb}
\usepackage{amsthm}
\usepackage{graphicx}
\input{Qcircuit}
\usepackage{bm}

\newcommand{\Fn}{\mathcal{F}}
\newcommand{\be}{\begin{eqnarray} \begin{aligned}}
\newcommand{\ee}{\end{aligned} \end{eqnarray} }
\newcommand{\benn}{\begin{eqnarray*} \begin{aligned}}
\newcommand{\eenn}{\end{aligned} \end{eqnarray*} }

\newcommand*{\textfrac}[2]{{{#1}/{#2}}}

\newcommand{\good}{\mathcal{G}ood}
\newcommand{\bad}{\mathcal{B}ad}

\newcommand*{\Enc}{\mathsf{Enc}}
\newcommand*{\Syn}{\mathsf{Syn}}

\newcommand*{\cA}{\mathcal{A}} 
\newcommand*{\cB}{\mathcal{B}}
\newcommand*{\cC}{\mathcal{C}}
\newcommand*{\cE}{\mathcal{E}}
\newcommand*{\cF}{\mathcal{F}}

\newcommand*{\cH}{\mathcal{H}}
\newcommand*{\cI}{\mathcal{I}}

\newcommand*{\cK}{\mathcal{K}}

\newcommand*{\cN}{\mathcal{N}}

\newcommand*{\cR}{\mathcal{R}}
\newcommand*{\cO}{\mathcal{O}}

\newcommand*{\cS}{\mathcal{S}}

\newcommand*{\cT}{\mathcal{T}}

\newcommand*{\cX}{\mathcal{X}}
\newcommand*{\cY}{\mathcal{Y}}
\newcommand*{\cZ}{\mathcal{Z}}

\newcommand{\iSet}{\mathcal{I}}

\newcommand*{\Ext}{\mathsf{Ext}}

\newcommand*{\tr}{\mathsf{tr}}
\newcommand*{\sbin}{\{0,1\}}

\newcommand*{\ExpE}{\mathop{\mathbb{E}}}
\newcounter{protoCount}
\newcounter{protoList}
\newsavebox{\tmpbox}
\newlength{\protobox}
\newenvironment{protocol}[3]{
\bigskip
\addtocounter{protoCount}{1}
\noindent \begin{lrbox}{\tmpbox}
\setlength{\protobox}{\textwidth/2-2.5ex}
\addtolength{\protobox}{-0.5cm}
\begin{minipage}[c]{\protobox}
\begin{bfseries}Protocol #1: #2\end{bfseries}
\ifthenelse{\equal{#3}{\empty}}{}{\\ #3}
\begin{list}{\begin{bfseries}\arabic{protoList}:\end{bfseries}}
{\usecounter{protoList}}
}{
\end{list}
\end{minipage}\end{lrbox}
\fbox{\usebox{\tmpbox}}
\bigskip
}

\newcommand{\bc}{\begin{center}}
\newcommand{\ec}{\end{center}}

\newcommand{\id}{\mathbb{I}}

\newtheorem{theorem}{Theorem}[section]

\newtheorem{lemma}[theorem]{Lemma}

\newtheorem{definition}[theorem]{Definition}

\newtheorem{corollary}[theorem]{Corollary}

\newcommand{\hil}{\mathcal{H}}

\newcommand{\hmin}{\ensuremath{{\rm H}_{\infty}}}
\newcommand{\hzero}{\ensuremath{{\rm H}_{0}}}
\newcommand{\hmineps}{\hmin^{\varepsilon}}

\newcommand{\textmath}{}

\usepackage{amsfonts}

\def\Complex{\mathbb{C}}

\def\id{\mathbb{I}}

\def\01{\{0,1\}}

\newcommand{\eps}{\varepsilon}

\newcommand{\proj}[1]{|#1\rangle\langle#1|}
\newcommand{\ketbra}[2]{|#1\rangle\langle#2|}

\newcommand{\rank}{\operatorname{rank}}

\newcommand{\bop}{\mathcal{B}}

\newcommand{\hout}{\mathcal{H}_{\rm out}}
\newcommand{\hin}{\mathcal{H}_{\rm in}}

\newcommand{\assign}{\ensuremath{\kern.5ex\raisebox{.1ex}{\mbox{\rm:}}\kern -.3em =}}

\newcommand{\nsw}{noisy-storage model}
\newcommand{\bsm}{bounded-storage model}
\newcommand{\scp}{strong-converse property}

\begin{document}
\title{Unconditional security from noisy quantum storage}
\author{Robert K{\"onig}$^{1,2}$, Stephanie Wehner$^{1,3}$, and J{\"u}rg Wullschleger$^{4,5}$\\
\normalsize{\textit{$^1$ Institute for Quantum Information, Caltech, Pasadena CA 91125, USA}}\\
\normalsize{\textit{$^2$ IBM Watson Research Center, Yorktown Heights, NY 10598, USA}}\\
\normalsize{\textit{$^3$ Centre for Quantum Technologies, National University of Singapore, 117543 Singapore}}\\
\normalsize{\textit{$^4$ Universit\'e de Montr\'eal, D\'epartement IRO, Montr\'eal (Qu\'ebec), H3C 3J7, Canada}}\\
\normalsize{\textit{$^5$ McGill University, School of Computer Science, Montr\'eal (Qu\'ebec), H3A 2A7, Canada}}
}
\maketitle

\begin{abstract}
We consider the implementation of two-party cryptographic primitives based on the sole assumption that no large-scale reliable quantum storage is available to the cheating party. We construct novel protocols for oblivious transfer and bit commitment, and prove that realistic noise levels provide security even 
against the most general attack. Such unconditional results were previously only known in the so-called \bsm\ which 
is a special case of our setting. Our protocols can be implemented with present-day hardware used for quantum key distribution. In particular, no quantum storage is required for the honest parties. 
\end{abstract}

\section{The \nsw: definition and results}
\subsection{Motivation: security from {\em physical  }assumptions} 
The security of most cryptographic systems currently in use is based on the premise that a certain computational problem is hard to solve for the adversary. Concretely, this relies on the assumption  that the adversary's computational resources are limited, and the underlying problem is hard in some precise complexity-theoretic sense.  While the former assumption may be justified in practice, the latter statement is usually an unproven mathematical conjecture. In contrast, quantum cryptographic schemes are designed in such a way that they provide security based solely on the validity of quantum physics. No assumptions on the adversary's computational power nor the validity of some complexity-theoretic statements are needed. 

Unfortunately, not even the laws of quantum physics allow us to realize all desirable cryptographic functionalities without further 
assumptions~\cite{lo:insecurity,mayers:trouble,lo&chau:bitcom2,lo&chau:bitcom,mayers:bitcom}. An example of such a functionality is (fully randomized) oblivious transfer, where Alice receives two random strings $S_0,S_1$, while Bob receives one of the strings $S_C$ together with the index~$C$. Security for this primitive means that neither Alice nor Bob can obtain any information beyond this specification. A protocol which securely implements oblivious transfer is desirable because any two-party computation, such as secure identification, can be based on this building block~\cite{kilian:foundingOnOT,GV87}.

In light of this state of affairs, it is natural to consider other {\em physical} assumptions: Motivated by similar classical models~\cite{Maurer90a,Maurer92b}, 
the authors of~\cite{serge:bounded,serge:new} and~\cite{steph:diss,prl:noisy,noisy:robust} propose to assume that the adversary's {\em quantum} storage is 
bounded and noisy, respectively. The assumption of bounded quantum 
storage deals with the noiseless case (but assumes a small amount of storage), whereas
the \nsw\ deals with the case of noise (but possibly a large amount of storage).
Here, we introduce a more general point of view which incorporates both the amount of storage and noise. 
We refer to this simply as the \emph{noisy-storage model}. The previously considered settings are special cases, as we will explain below.

Compared to the classical world, the assumption of limited and noisy quantum storage is particularly realistic in view of the 
present state of the art, and the 
considerable challenges faced when trying to build scalable quantum memories. 
Indeed, it is unknown whether it is physically possible to build noise free memories.
Further motivation for considering noise as a resource for security over the mere assumption of bounded storage
comes from the fact that the transfer of the state of a (photonic) qubit used during the execution of the 
protocol onto a different carrier used as a quantum memory 
(such as an atomic ensemble) is typically already noisy. 

\subsection{Contribution and methods}
We consider the noisy-storage model which was previously introduced in~\cite{steph:diss,prl:noisy,noisy:robust} where it appeared in a slightly more specialized form.
All previous security proofs in this model required additional assumptions beyond having noise. 
In particular, in the analysis of~\cite{prl:noisy}, the adversary was restricted to performing individual attacks using product measurements on the qubits received in the 
protocol. This is a significant restriction as multi-qubit measurements are possible even today, and can be compared to an analysis of quantum key distribution~\cite{bb84}
 where the eavesdropper is restricted to measuring each qubit individually.
We provide a fully general proof of security against arbitrary attacks that bit-commitment and oblivious transfer can be achieved
in the general noisy-storage model. This encompasses and extends all previously considered settings~\cite{serge:bounded,serge:new,steph:diss,prl:noisy,noisy:robust}. 
As a side effect, we also obtain significantly improved parameters for the special case of bounded storage.

In order to obtain this result, we require a number of methods that have not been used before either
in the noisy- or bounded-quantum-storage setting.
\begin{itemize}
\item We formally relate the security of our protocols to the problem of sending information through the noisy-storage channel.
This is very intuitive, and much more natural than previous approaches such as the restriction to individual attacks in the noisy-storage
model~\cite{prl:noisy}, or the assumption of bounded storage~\cite{serge:bounded}. More specifically, we show that a sufficient condition
for security is that the number of classical bits that can be sent through the noisy-storage channel is limited.
We introduce our generalized model in Section~\ref{sec:nsw}, and state our result in Section~\ref{sec:mainresult}.

\item We introduce a novel cryptographic primitive called weak string erasure (see Section~\ref{sec:wse}) that may be of independent interest.
We provide a simple quantum protocol that securely realizes weak string erasure in the noisy-storage model, in which the honest parties
do not require any quantum memory at all to execute the protocol. Our protocol can be implemented with present-day technology.
In our security proof, we require information-theoretic tools such as the recently proven strong converse for channel coding~\cite{rs:converse}.

\item We construct new protocols for bit commitment and oblivious transfer based on weak string erasure, and prove security against arbitrary 
attacks. Our protocols are purely classical, merely using the simple quantum primitive of weak string erasure which is a conceptually appealing feature. 
We make use of various techniques such as error-correcting codes, privacy amplification, interactive hashing and min-entropy sampling
with respect to a quantum adversary.

\end{itemize}
Our work raises many immediate open questions and has already led to follow-up work which we discuss in Section~\ref{sec:conclude}.
 
\subsection{The \nsw\ }\label{sec:nsw}

Let us now describe more formally what we mean by a noisy quantum memory. We think of a device whose input states are in some Hilbert space~$\cH_{in}$. A state $\rho$ stored in the device decoheres over time. That is, the content of the memory after some time~$t$ is a state $\cF_t(\rho)$, where $\cF_t:\cB(\cH_{in})\rightarrow\cB(\cH_{out})$ is a completely positive trace-preserving map corresponding to the noise in the memory. 
Since the amount of noise may of course depend on the storage time, the behaviour of the storage is completely described by the family 
of maps~$\{\cF_t\}_{t>0}$.
We will make the minimal assumption that the noise is Markovian,
that is, the family $\{\cF_t\}_{t>0}$ is a continuous one-parameter semigroup
\begin{align} 
\cF_0=\id\qquad\textrm{ and }\qquad \cF_{t_1+t_2}=\cF_{t_1}\circ\cF_{t_2}\ .\label{eq:markovproperty} 
\end{align} 
This tells us that the noise in storage only increases with time, and
is essential to ensure that the adversary cannot gain any information by delaying the readout~\footnote{This property is implicitly assumed in~\cite{prl:noisy}.}. 
This is the only restriction imposed on the adversary who may otherwise be all-powerful. 
In particular, we allow that all his actions are instantaneous, including computation, communication, measurement and state preparation.

How can we hope to obtain security in such a model? In our protocol, 
we will introduce certain time delays $\Delta t$ which force any adversary to use his storage device for a time at least $\Delta t$. 
Our assumptions imply that the best an adversary can do is to read out the information from the device immediately after 
time~$\Delta t$, as any further delay will only degrade his information further. We can thus focus on the channel $\cF=\cF_{\Delta t}$ when analyzing security
instead of the family $\{\cF_t\}_{t\geq 0}$. 
Note that since the adversary's actions are assumed to be instantaneous, he can use any error-correcting code even if the best encoding and decoding procedure may be difficult to perform.
Summarizing, our model assumes that
\begin{itemize}
\item
The adversary has unlimited classical storage, and (quantum) computational resources.
\item
Whenever the protocol requires the adversary to wait for a time $\Delta t$, he has to measure/discard all his quantum information except what he can encode (arbitrarily) into $\cH_{in}$. This information then undergoes noise described by~$\cF$.
\end{itemize}
To see how previously analyzed cases fit into our model, note that the \bsm\ corresponds to the case where $\cH_{in}$ is of limited input dimension, and $\cF$ is the identity on $\cH_{in}.$ 
Concretely,~\cite{serge:new} considers protocols with $n$~qubits of communication and $\cH_{in}\cong(\mathbb{C}^2)^{\otimes \nu n}$ for some parameter~$\nu>0$ which we call the {\em storage rate}. Security of certain protocols was established for $\nu < 1/4$. Furthermore, the protocol proposed by Cr{\'e}peau~\cite{crepeau:qot} for oblivious transfer is secure if the adversary cannot store any quantum information at all, corresponding to a storage rate of $\nu = 0$.
Previous work on the \nsw\ ~\cite{prl:noisy} analyzed
protocols with $n$~qubits of communication, where 
the noise $\cF\equiv \cN^{\otimes n}$ is an $n$-fold tensor product of a noisy single-qubit channel 
$\cN:\cB(\mathbb{C}^2)\rightarrow\cB(\mathbb{C}^2)$ (i.e., $\cH_{in}\cong(\mathbb{C}^2)^{\otimes n}$ and $\nu=1$). Note, however, that in~\cite{prl:noisy} the adversary was further restricted to performing product measurements
on the qubits received in the protocol (albeit otherwise fully arbitrary).

\subsection{Main result}\label{sec:mainresult}

We now state our main result of establishing security in the \nsw\ against \emph{fully general attacks} for arbitrary channels $\cF:\cB(\cH_{in})\rightarrow\cB(\cH_{out})$.
As explained, we form a very natural relation between the 
security of our protocols and the problem of transmitting information through the noisy-storage 
channel\footnote{The 
communication problem is equivalent to storing the string, and later trying to read it from the device.}.
More specifically, we prove that a sufficient condition for security is that the number of classical bits that can be sent through the 
noisy storage-channel is limited. 

As usual in cryptography, we would like to compare the adversary's resources to those of the honest parties and/or the complexity of operations used in the protocol. Here we parametrize these by the number $n$~of qubits transmitted during the protocol. For the adversary's storage, we therefore consider a family $\{\Fn\}_n$ of storage devices. The quality of the adversary's storage can then be measured (for a fixed $n$) by the following operational quantity: the success probability of correctly transmitting
a randomly chosen $n R$-bit string
$x \in \sbin^{nR}$ through the storage device~$\Fn$, which can be written as
\begin{align}
P_{succ}^{\Fn}(nR):=\max_{\{D_x\}_x,\{\rho_x\}_x}\frac{1}{2^{nR}}\sum_{x\in\sbin^{nR}} \tr(D_x\Fn(\rho_x))\ , \label{eq:succ}
\end{align}
where the maximum is taken over families of code states $\{\rho_x\}_{x\in\sbin^{nR}}$ on $\cH_{in}$ and decoding POVMs $\{D_x\}_{x\in\sbin^{nR}}$ on~$\cH_{out}$. 
We show that security can be obtained for arbitrary channels with the property that the decoding probability decays exponentially above a certain threshold:

\begin{theorem}[Informal statement]\label{thm:maininformal}
Suppose that for the family of channels $\{\Fn\}_n$
and the constant $0 < R < 1/2$ there exist constants $n_0 > 0$ and $\gamma > 0$ such that for all $n \geq n_0$ 
the decoding probability satisfies
\begin{align}
P^{\Fn}_{\rm succ}(nR) \leq 2^{- \gamma n}\label{eq:expdecaypropgeneral}\ .
\end{align}
Then oblivious transfer and bit commitment can be implemented using $O(n)$~qubits of communication against an adversary whose noisy storage is described by the family $\{\Fn\}_n$.
Moreover, the security is exponential in~$n$. 
\end{theorem}
Remarkably, the statement of Theorem~\ref{thm:maininformal} does not require any knowledge of the channel $\Fn$
beyond its relation to the coding problem. In particular, we do not need to assume that $\Fn$ is of tensor product form. 
This includes for example the practically interesting case where errors are likely to occur in bursts in the storage device, or the
noisy channel itself has memory.
We discuss possible extensions
and limitations of our approach in Section~\ref{sec:extensions}. 
We point out that the length of the input strings used in oblivious transfer and bit commitment 
per communicated qubit depends on the exponent~$\gamma$ in~\eqref{eq:expdecaypropgeneral}; this is hidden in the constant in the $O$-expression in Theorem~\ref{thm:maininformal}.

Determining the constant $\gamma$ is of course no easy task for arbitrary storage devices. 
To obtain explicit security parameters, we thus proceed to 
consider the special case where the channels
are of the form $\Fn = \cN^{\otimes \nu n}$ 
where $n$~is the number of qubits sent in the protocol, and $\nu\geq 0$ is the \emph{storage rate}.  
Our proof connects the security of protocols in the \nsw\ for such channels to the \emph{classical capacity}~$C_\cN$ of~$\cN$. 
This provides a quantitative expression of our intuition that 
noisy channels which are of little use for classical information transmission give rise to security in the \nsw.
First of all, observe that there can only exist a constant $\gamma > 0 $ leading to the exponential decay of~\eqref{eq:expdecaypropgeneral}
if the {\em classical capacity} $C_{\cN}$ of the channel is strictly smaller than the rate $R$ at which we send information through the channel. 
This, however, is not sufficient, since $R > C_{\cN}$ is not generally known to imply~\eqref{eq:expdecaypropgeneral} for $\Fn=\cN^{\otimes n}$. We are therefore interested in channels~$\cN$ which satisfy the following \emph{\scp}: The success probability~\eqref{eq:succ} decays exponentially for rates $R$ above the capacity, i.e., it takes the form
\begin{align}
\begin{split}
P_{succ}^{\cN^{ \otimes n}}(nR)&\leq 2^{-n\gamma^\cN(R)}\qquad\textrm{ where}\\
 \gamma^\cN(R)&>0\qquad\textrm{ for all }\qquad R>C_\cN .
\end{split}\label{eq:strongconverseproperty} 
\end{align} 
In~\cite{rs:converse}, property~\eqref{eq:strongconverseproperty} was shown to hold for a large class\footnote{The result of~\cite{rs:converse} applies to channels with certain covariance properties and additive minimum output $\alpha$-R\'enyi entropy. Examples are all unital qubit channels, the Werner-Holevo channel and the depolarizing channel.} of channels, including the depolarizing channel (see~\eqref{eq:depolarizingchannel} below). It was also shown how to compute $\gamma^\cN(R)$.
Combining Theorem~\ref{thm:maininformal} with~\eqref{eq:strongconverseproperty}, we obtain the following statement:
\begin{corollary}[Informal statement]\label{cor:mainstatement} 
Let $\nu\geq 0$, and suppose that $\cN$ satisfies the \scp~\eqref{eq:strongconverseproperty}. If \begin{align*}
C_\cN\cdot  \nu < \frac{1}{2}\ ,
\end{align*}
then oblivious transfer and bit commitment can be
implemented with polynomial resources (in $n$) and exponential security against an adversary with noisy storage $\Fn = \cN^{\otimes \nu n}$.
For the special case of bounded (noise-free) qubit storage ($C_\cN = 1$) this gives security for $\nu < 1/2$.
\end{corollary}

An important example for which we obtain security is the $d$-dimensional depolarizing channel~$\cN_r:\cB(\mathbb{C}^d)\rightarrow\cB(\mathbb{C}^d)$ defined for $d \geq 2$ as
\begin{align}
\cN_r(\rho) := r \rho + (1-r) \frac{\id}{d}\ \textrm{ for some fixed } 
0 \leq r\leq 1\ , \label{eq:depolarizingchannel}
\end{align}
which replaces the input state $\rho$ with the completely mixed state with probability~$1-r$. For $d=2$, this means that the adversary
can store $\nu n$~qubits, which are affected by independent and identically distributed noise. It has been shown
that the depolarizing channel exhibits the strong-converse property~\cite{rs:converse}.
To see for which values of $r$ we can obtain security, we need to consider the classical capacity of the depolarizing channel as 
evaluated by King~\cite{king:depol}. For $d=2$, i.e., qubits, it is given by
\begin{align*}
C_{\cN_{r}}=1+\frac{1+r}{2}\log\frac{1+r}{2}+\frac{1-r}{2}\log\frac{1-r}{2}\ . 
\end{align*}
\begin{figure}[t!]
\hspace{-5ex}\scalebox{0.85}{
\begin{pspicture}(-1.0,0)(12.0,5.0)
\psset{unit=.7cm}
\psset{linewidth=.8pt}
\psset{labelsep=2.5pt}
\put(0,0){\epsfig{file=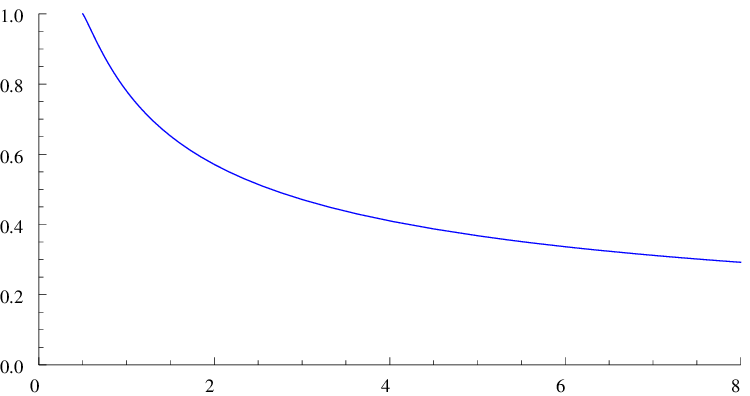,width=10cm}}
\put(2,2){secure realization}
\put(8,6){?}
\put(0.7,7.5){$r$}
\put(14.3,0.6){$\nu$}
\end{pspicture}
}
\caption{Our results applied to depolarizing noise $\cF=\cN_r^{\otimes \nu n}$: The vertical axis represents the noise parameter $r$, while the horizontal axis represents the storage rate $\nu$.
Our protocols are secure when the pair $(r,\nu)$ is in the lower region bounded by the solid blue curve.
Security is still possible in the region labeled with '?', but cannot be obtained from our analysis.
\label{fig:depolarizingchannelanalysis}}
\end{figure}
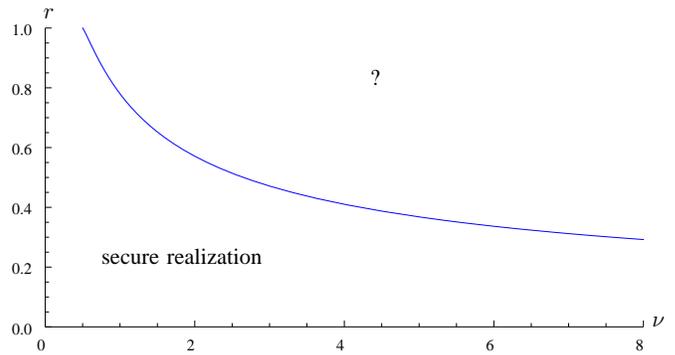 

Figure~\ref{fig:depolarizingchannelanalysis} shows the region in the $(r,\nu)$-plane corresponding to the noise channel~$\cF=\cN_r^{\otimes\nu n}$, where we allow $n$~qubits of communication in the protocol. This is obtained from Corollary~\ref{cor:mainstatement} (The depolarizing channel~$\cN_r$ satisfies the corresponding conditions).

\subsubsection*{Comparison to the \bsm: depolarizing noise}
It was previously observed~\cite{chris:diss} that the case of 
depolarizing storage noise (i.e., $r<1$) can be dealt with using results obtained in the \bsm\  (i.e., $r=1$) when the noise is 
sufficiently strong. More precisely, the results of~\cite{serge:new} can be extended to give non-trivial statements if the ``effective'' dimension of the storage system to be less than~$n/4$, where~$n$ is the number of qubits communicated in the 
protocol~\footnote{We compare the randomized oblivious transfer protocol of~\cite{serge:new} to our protocol
based on weak string erasure.}.
We sketch such a simple dimensional analysis
to illustrate that our model offers significant improvements over the bounded-storage analysis: we obtain security even at lower noise levels and higher storage rates. 

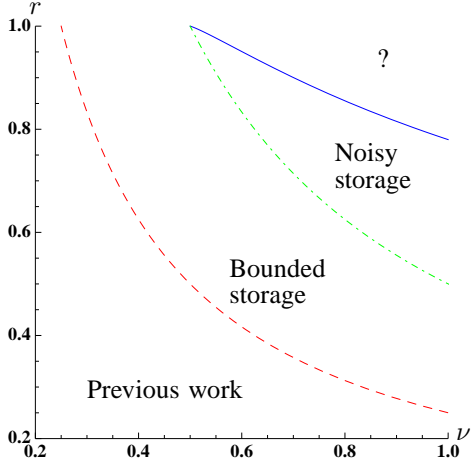
\begin{figure}[h]
\begin{center}
\scalebox{1.0}{
\begin{pspicture}(-1.0,0)(6.0,6.0)
\psset{unit=.7cm}
\psset{linewidth=.8pt}
\psset{labelsep=2.5pt}
\put(0,0){\epsfig{file=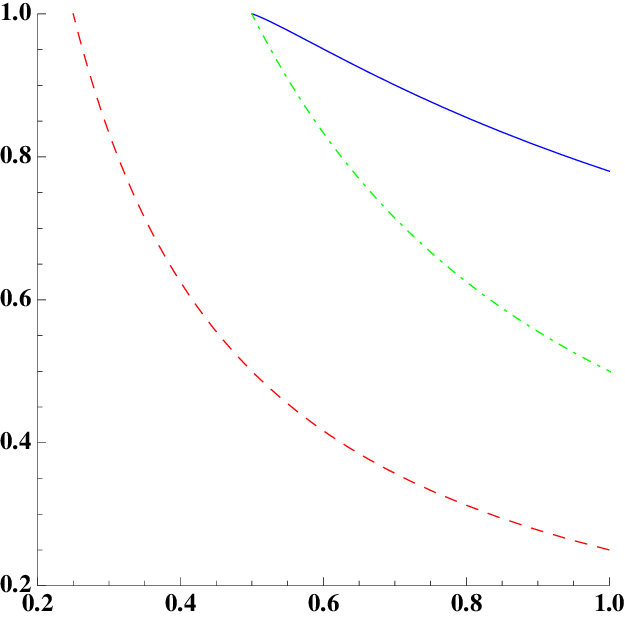,width=6cm}}
\put(1.5,1.2){Previous work}
\put(4.2,3.5){Bounded}
\put(4.2,3){storage}
\put(6.2,5.7){Noisy}
\put(6.2,5.2){storage}
\put(7,7.5){?}
\put(0.4,8.5){$r$}
\put(8.5,0.4){$\nu$}
\end{pspicture}\hspace{2.8cm} 
}
\end{center}
\caption{Security for depolarizing noise parameters $(1,\nu)$ with $\nu<1/4$ was established in the \bsm\ (BSM). 
Our simple argument 
Our more refined protocols and analysis give significantly improved parameters of $\nu < 1/2$ for the \bsm, for which the same argument extends security to the region
bounded by the green dot-dashed curve. However, our work still extends this 
region even further by considering noisy instead of merely bounded storage (solid blue curve).  
We stress that such a na{\"i}ve dimensional analysis does not apply to other channels (such as the Two-Pauli channel), 
while our more refined analysis gives results even in such cases. 
\label{fig:comparison}}
\end{figure}

Concretely, consider the noise channel~$\cF=\cN_r^{\otimes \nu n}:\cB((\mathbb{C}^2)^{\otimes \nu n})\rightarrow\cB((\mathbb{C}^2)^{\otimes \nu n})$ (cf.~\eqref{eq:depolarizingchannel} for~$d=2$).
 Applying depolarizing noise to any of the $\nu n$ systems $\Complex^2$ means
that the state on this system  is replaced by the completely mixed state with probability~$1-r$.
We can think of an indicator random variable $E^{\nu n}=(E_1,\ldots,E_{\nu n})\in\sbin^{\nu n}$, where~$E_i$ is~$1$ if and only if the 
$i$-th qubit is replaced by the completely mixed state. These ``erasure'' variables are independent and identically distributed Bernoulli variables with parameter~$r=P_{E_i}(0)$.  In particular, the number of erasures 
\begin{align*}
|E^{\nu n}|=\sum_{i=1}^{\nu n} E_{i}
\end{align*}
is distributed according to the binomial distribution with $\nu n$ trials, each of which succeeds with probability $1-r$. 

We now assume that the adversary is given the location of the erasures~$E^{\nu n}$ in addition to the output of the channel. Note that this can only make the adversary more powerful. 
Conditioned on the locations $E^{\nu n}$, the ``effective dimension'' of his channel is equal to~$2^{\nu n-|E^{\nu n}|}$. 
Hence, we may think of an ``effective'' storage rate $\nu_{\textrm{eff}}$ given by the random variable
\begin{align*}
\nu_{\textrm{eff}} &=\nu-\frac{|E^{\nu n}|}{n}\ . 
\end{align*}
We know from the bounded storage model analysis~\cite{serge:new} that for $\nu_{\textrm{eff}}<\frac{1}{4}$, the previously studied protocols provide security. 
Overall, we therefore conclude that security can be obtained from the noisy channel~$\cF$ if~$\Pr[\nu_{\textrm{eff}} > \frac{1}{4}]$ is exponentially small. Note that by Chernoff's inequality
\begin{align*}
\begin{split}
\Pr\left[\nu_{\textrm{eff}}* > 1/4\right]=\Pr\left[|E^{\nu n}|<(1-\delta)\mu\right]<e^{-\mu \delta^2/2}\\
\textrm{if }\delta=\frac{1-4\nu r}{4\nu (1-r)}>0\ ,\ \textrm{ where }\mu=n\nu (1-r)\ .
\end{split}
\end{align*}
In particular, we conclude that we obtain security for
\begin{align}
\nu r<\frac{1}{4}\ .\label{eq:bsmanalysis}
\end{align}
Figure~\ref{fig:comparison} compares the curve of this equation~\eqref{eq:bsmanalysis} to the results we will derive below.
We see that for the noiseless case ($r=1$), our analysis provides security for storage rates $\nu < 1/2$, extending previous results (i.e., $\nu<1/4$ in~\cite{serge:new}) in the \bsm. This improvement stems from the fact that 
(for oblivious transfer) our protocol uses a different classical post-processing based on 
interactive hashing instead of the min-entropy splitting tool of~\cite{serge:new}. Note that this requires additional rounds of
classical communication.

One may wonder whether a security proof may alternatively be obtained based on the idea of simulating the storage noise~$\Fn=\cN^{\otimes\nu n}$ using a limited number of qubits.
For channels without memory, the quantum 
reverse Shannon theorem~\cite{andreas:qrst} tells us that $\Fn$ can  be simulated 
using a certain number of (noise-free) qubits when the sender and receiver share entanglement. 
Hence the total size of the system consisting of the noise-free qubits and the entanglement is rather large.
However, as explained in~\cite{andreas:qrst}, the theorem implies an exponential decay of the decoding probability as in~\eqref{eq:strongconverseproperty}, but only for rates $R$~greater than the {\em entanglement-assisted capacity} of the channel~$\cN$.
Our security results thus extend to this regime by our new analysis. 
The fact that the entanglement-assisted capacity is generally greater than the unassisted one suggests that such a simulation-based approach is suboptimal: we are essentially overestimating the adversary's capabilities by allowing him to use (noise-free, time-like) entanglement. 

Let us give a 
simple concrete example that provides some intuition on why bounding the adversary's information by the size of his storage device is generally undesirable. Imagine that the adversary's channel $\Fn$ replaces the $n$~input qubits by a fixed state with overwhelmingly high probability and leaves the input untouched with negligible probability~$2^{-n}$. Clearly, the number of noise-free qubits required to simulate this channel is equal to~$n$, yet the adversary's decoding probability will be exponentially small. Simply bounding the adversary's information gain in terms of his storage as in the bounded-storage analysis~\cite{serge:new} therefore significantly overestimates his abilities.\footnote{One may argue that this example is artificial, and can easily dealt with by ``smoothing'': the channel is exponentially close to one which can be simulated with no qubits at all. More complex examples exist even classically: Imagine the adversary has some information
$B$ about a randomly chosen $n$-bit classical string $X$, where $B$ is the result of sending the string $X$ through a classical channel that
outputs the first $i \in \{1,\ldots,n\}$ bits of $X$ with probability $p_i = 2^{-i}$ for $i < n$ and $p_i = 2 \cdot 2^{-i}$ for $i = n$.
In cryptography the adversary's information is measured in terms of the min-entropy
$\hmin(X|B)=n - \log(1/(n+1))$. 
Furthermore, $\hzero(B) = \log \rank(B) = n$ and for even for a small smoothing parameter~$\eps= 2^{-k}$, one still has $\hzero^{\eps}(B) \geq n - \log(1/\eps)$. 
Knowing the size of $B$ only gives us the trivial bound
$\hmin(X|B) \geq \hmin(X) - \hzero(B) = 0$, although the conditional min-entropy is almost maximal.}

\subsection{Techniques: weak string erasure}
Before describing our protocols and proving Theorem~\ref{thm:maininformal}, we give a short overview of the techniques involved.

First, we introduce a primitive called {\em weak string erasure}, which may be of independent interest.
Our protocols for oblivious transfer and bit commitment are then based on this primitive.
Weak string erasure provides Alice with a random bit-string $X^n\in\sbin^n$, while Bob receives a randomly chosen substring~$X_{\iSet}=(X_{i_1},\ldots,X_{i_r})$, together with the index set $\iSet=\{i_1,\ldots,i_r\}$ specifying the location of these bits. 
Security of weak string erasure roughly means that Bob will remain ignorant about 
a significant amount of information about~$X^n$, while 
security against Alice means that she does not learn anything about $\iSet$ (for a precise definition, we defer the reader to Section~\ref{sec:wse}).

We provide a protocol for weak string erasure in the \nsw. This protocol can be implemented with present-day hardware used for quantum key distribution. In particular, it does not require the honest parties
to have any form of quantum memory. We  prove security  of this protocol for channels~$\cF$ as stated in Theorem~\ref{thm:maininformal}. Security against (even an all-powerful) Alice follows from the fact that the protocol only involves one-way communication from Alice to Bob. The security analysis in the presence of a malicious Bob limited by storage noise~$\cF$  is more involved. Our proof combines an entropic uncertainty relation involving post-measurement 
information~\cite{ww:pistar,serge:new} with a reformulation of the problem as a coding scheme: Essentially, the uncertainty relation implies that with high probability (over measurement outcomes),  Bob's {\em classical} information  about~$X^n$ before using his storage is limited. We then show that this implies 
that any successful attacker Bob needs to encode classical information at a high rate into his storage device. 
However, the assumed noisiness of~$\cF$ precludes this.

Having built a protocol for weak string erasure, we proceed to present protocols for bit commitment and oblivious transfer. The case of bit commitment is particularly appealing: It is essentially only based on weak string erasure and a classical code, and requires little additional analysis. Our approach to realizing oblivious transfer is somewhat more involved: Here weak string erasure is combined with 
a technique called {\em interactive hashing}~\cite{savvides:diss}. 
The output of interactive hashing is a pair of substrings of~$X$, one of which is completely known to Bob, while 
he only has partial knowledge about the other.  Privacy amplification~\cite{renato:compose} is then used to extract completely random bits. 
The security analysis of this protocol requires the use of entropy sampling with respect to a a quantum adversary~\cite{kr:sampling}.

As as side remark, note that Kilian~\cite{kilian:foundingOnOT} showed that oblivious transfer is 
universal for secure two-party computation. In particular, bit commitment could be built from oblivious transfer, but this reduction is generally inefficient. 

\section{Tools}\label{sec:prelim}
We briefly introduce all necessary notation as well as several important concepts we will need throughout the paper. For weak string erasure we require
the notion of min-entropy (Section~\ref{sec:minentropy}), uncertainty relations (Section~\ref{sec:uncertainty}), as well as an understanding
of how storage noise leads to information loss for the cheating party (Section~\ref{sec:infoLoss}). In our protocols for bit-commitment and oblivious transfer from weak string erasure,
we additionally require the concepts of smooth min-entropy (Section~\ref{sec:smoothminentropy}) and 
secure keys (Section~\ref{sec:nonuniformity}) respectively, and
 a number of tools, namely privacy amplification (Section~\ref{sec:privacyamplification}), sampling of min-entropy (Section~\ref{sec:minentropysampling}), and
finally interactive hashing (Section~\ref{sec:interactivehashing}).

\subsection{Notation}
For an integer~$n$, let $[n] \assign \{1,\ldots,n\}$. We use $2^{[n]} \assign \{\cS\ |\ \cS\subseteq [n]\}$ to refer to the set of all possible subsets of~$[n]$, including the empty set~$\emptyset$. For an $n$-tuple $x^n=(x_1,\ldots,x_n)\in\cX^n$ over a set $\cX$ and a (non-empty) set $\cI=\{i_1,\ldots,i_\ell\}\in 2^{[n]}$, we write~$x_{\cI}$ for the subtuple $x_{\cI}=(x_{i_1},\ldots,x_{i_\ell})\in\cX^\ell$. 

We use upper case letters to denote a random variable $X$ distributed according to a distribution~$P_X$ over a set~$\cX$, 
and use lower case letters $x$ for elements~$x \in \cX$. Joint distributions of~e.g., three random variables $(X,Y,Z)$ on $\cX\times\cY\times\cZ$ are denoted by~$P_{XYZ}$. Given a function $f:\cX\rightarrow\cY$, any distribution $P_{X}$ of  a random variable~$X$ gives rise to another jointly distributed random variable $Y=f(X)$: The joint distribution $P_{XY}\equiv P_{Xf(X)}$ is given by
\begin{align}
P_{Xf(X)}(x,y)&=P_{X}(x)\delta_{f(x),y}\ ,\label{eq:functionappliedprob}
\end{align} 
where $\delta_{i,j}$ is the Kronecker symbol. An important example is the case where~$X^n \in \sbin^n$ is a random bitstring and $\cI \in 2^{[n]}$ is a random subset of $[n]$, where $X^n$ and $\cI$ have joint distribution $P_{X^n\cI}$. In this case, the joint distribution  $P_{X^n\cI Z}\equiv P_{X^n\cI X_{\cI}}$ describes e.g., a situation where some bits $Z = X_{\cI}$ of a string $X^n$ are given, together with a specification $\cI$ 
of where these bits are located in the original string.

We use $\bop(\hil)$ to denote the set of bounded operators on a Hilbert space $\hil$. A (quantum) \emph{state} is 
a Hermitian operator
$\rho \in \bop(\hil)$ satisfying $\tr(\rho) = 1$ and $\rho \geq 0$. Quantum states can be used to encode classical probability distributions: for a (finite) set $\cX$, we fix a Hilbert space $\cH_{\cX}\cong (\mathbb{C}^{|\cX|})$ and an orthonormal basis $\{\ket{x}\ | \ x\in\cX\}$ of $\cH_{\cX}$. This will be referred to as the {\em computational basis}. A probability distribution~$P_X$ on $\cX$ can then be encoded into the classical state ($c$-state)
$$
\rho_X = \sum_{x \in \cX} P_X(x) \proj{x}\ .
$$
Of particular interest is the uniform distribution over $\cX$, which gives rise to the completely mixed state on $\cH_\cX$ denoted 
by the shorthand
$$
\tau_{\cX} \assign \frac{1}{|\cX|} \sum_{x \in \cX} \proj{x}\ .
$$
States describing classical information (random variables) and truly quantum information simultaneously are termed {\em classical-quantum} or cq-states. They are described by bipartite systems, where the classical part of the state is diagonal with respect to the computational basis.  Concretely, let $\cH_Q$ be an additional Hilbert space. A state $\rho_{XQ}$ on $\cH_\cX\otimes\cH_Q$ is a cq-state if it has the form
\begin{align}
\rho_{XQ}=\sum_{x\in\cX} P_X(x)\underbrace{\proj{x}}_{X}\otimes\underbrace{\rho_x}_Q\ .\label{eq:cqstate}
\end{align}
In other words, such  a state $\rho_{XQ}$ encodes an ensemble of states $\{P_X(x),\rho_x\}_{x\in\cX}$ on $\cH_Q$, where $\rho_x$ is the {\em conditional state on $Q$ given $X=x$}. The notion of cq-states directly generalizes to multipartite systems, where classical parts are diagonal with respect to the computational basis. We often fix an ordering of the multipartite parts, and indicate by $c$ or $q$ whether 
a part is classical or quantum. We can also apply functions to classical parts as before. For a function $f:\cX\rightarrow\cY$,
\begin{align}
\rho_{Xf(X)Q}&=\sum_{\substack{x \in \cX,\\ y \in \cY}}P_{Xf(X)}(x,y)\underbrace{\proj{x}}_X\otimes\underbrace{\proj{y}}_{f(X)}\otimes\underbrace{\rho_x}_{Q}\ \label{eq:xfxq}
\end{align}
is the ccq-state encoding the pair $(X,f(X))$ of classical random variables (cc) distributed according to~\eqref{eq:functionappliedprob} as well as the quantum information~$Q$ (q) (which depends only on $X$ in this case). Note that in~\eqref{eq:xfxq}, the systems on the rhs.~are uniquely determined by the expression on the lhs. We will therefore omit the braces below. Given a state $\rho_{Q_1 Q_2}$ on systems $Q_1$ and $Q_2$, we also 
use $\rho_{Q_1} = \tr_{Q_2}(\rho_{Q_1 Q_2})$ to denote the state obtained by tracing out $Q_2$. 

The {\em Hadamard transform} is the unitary described by the matrix~
$$
H=\frac{1}{\sqrt{2}}\left(\begin{matrix}1 & 1\\
1&-1\end{matrix}\right)
$$ 
in the computational basis $\{\ket{0},\ket{1}\}$ of the qubit Hilbert  space~$\mathbb{C}^2$.
For the $n$-qubit Hilbert space, we let
\begin{align*}
\begin{split}
H^{\theta^n}\ket{x^n}&:= H^{\theta_1}\ket{x_1} \otimes \ldots \otimes H^{\theta_n}\ket{x_n}\\
\mbox{ for } x^n &= (x_1,\ldots,x_n),\  \theta^n = (\theta_1,\ldots,\theta_n) \in \01^n\ .
\end{split}
\end{align*}
We also call states of this form {\em BB84-states}. 

Finally, we need a distance measure for quantum states on a Hilbert space~$\cH$. We use the distance determined by the trace norm
 $\|A\|_1:=\tr \sqrt{A^\dagger A}$ for bounded operators $A\in \cB(\cH)$. We will say 
that two states $\rho,\sigma\in\cB(\cH)$ are \emph{$\varepsilon$-close} if $\frac{1}{2}\|\rho-\sigma\|_1\leq \varepsilon$, which we also write
as
$$
\rho \approx_\eps \sigma\ .
$$

\subsection{Quantifying adversarial information}\label{sec:info}\label{sec:advinformation}
\subsubsection{Min-entropy and measurements}\label{sec:minentropy}
One of the main properties of the weak string erasure-primitive is that the adversary's (quantum) information~$Q$ about the generated bit-string~$X$ is limited. 
To make this statement precise, we first need to introduce an appropriate measure of information.
Throughout, we are interested in the case where the adversary holds some (possibly quantum) information~$Q$ about a classical random variable~$X$. This situation is described by a cq-state $\rho_{XQ}$ as in~\eqref{eq:cqstate}. A natural measure for the amount of information~$Q$ gives about~$X$ is the maximal average success probability that a party holding~$Q$ has in guessing
the value of~$X$. For a given $cq$-state $\rho_{XQ}$, this {\em guessing probability} can be written as
\begin{align}\label{eq:pguessdef}
P_{guess}(X|Q):=\max_{\{D_x\}_x}\sum_{x} P_X(x)\tr(D_x\rho_x)\  ,
\end{align}
where the maximization is over all POVMs~$\{D_x\}_{x\in\cX}$ on $\cH_Q$. 
It will be convenient to turn~\eqref{eq:pguessdef} into an conditional entropy-like quantity, called the \emph{min-entropy}, which is given
by~\footnote{All logarithms are taken to base~2.}
\begin{align}
\hmin(X|Q):=-\log P_{guess}(X|Q)\ .\label{eq:minentropyguess}
\end{align}
Note that the min-entropy was originally defined~\cite{renato:diss} for arbitrary bipartite states~$\rho_{AB}$, as we will discuss in more detail below.

As an illustrative, yet important, example consider the following ccq-state on $\cH_X\otimes\cH_\Theta\otimes\cH_Q\cong (\mathbb{C}^2)^{\otimes 3}$
\begin{align}
\rho_{X\Theta Q}&=
\frac{1}{4}\sum_{x,\theta \in \sbin} \proj{x}\otimes\proj{\theta}\otimes H^{\theta} \proj{x} H^{\theta}
\label{eq:entropyuncertaintystate}
\end{align}
This state arises when encoding a uniformly random bit $X$ using either the computational basis ($\Theta=0$) or the Hadamard basis ($\Theta=1$) chosen uniformly 
at random. Clearly, we have
\begin{align*}
\hmin(X)&=1\ ,\\ 
\hmin(X|\Theta)&=1\ ,\qquad\textrm{ and }\\
\hmin(X|Q\Theta)&=0\ ,
\end{align*}
where the last identity is a consequence of the fact that given $\Theta=\theta$, the operation $H^\theta$ can be undone, such that a subsequent 
measurement in the computational basis provides $X$ with certainty. Note that this is a special case of the identity
\begin{align}
\hmin(X|Q\Theta)&=-\log\ExpE_{\theta\leftarrow P_\Theta} \left[2^{-\hmin(X|Q,\Theta=\theta)}\right]\ ,\label{eq:classicalconditioning}
\end{align}
for a general cq-state $\rho_{XQ\Theta}$ with classical part~$\Theta$, where $\ExpE_{\theta\leftarrow P_\Theta}$ denotes the expectation value over the choice of~$\Theta$, and $\hmin(X|Q,\Theta=\theta)$ is the min-entropy of the conditional state
\begin{align*}
\rho_{X|Q,\Theta=\theta}&=\sum_{x \in \sbin} P_{X|\Theta=\theta}(x)\proj{x}\otimes H^{\theta} \proj{x} H^{\theta}\ .
\end{align*}
Returning to the state~\eqref{eq:entropyuncertaintystate}, it can also be shown~\cite{ww:pistar} that
\begin{align*}
\hmin(X|Q)= - \log \left( \frac{1}{2} + \frac{1}{2\sqrt{2}}\right)\ .
\end{align*}

\subsubsection{Smooth min-entropy}\label{sec:smoothminentropy}
When building oblivious transfer from weak string erasure, we will need to employ a more general definition of the min-entropy given in~\cite{renato:diss}.
For arbitrary (not necessarily unit-trace, or cq) bipartite density operators $\rho_{AB}$ this quantity is defined as
\begin{align}
\hmin(A|B)_{\rho}&=-\log\inf_{\substack{\sigma_B\geq 0\\ 
\rho_{AB}\leq \id_A\otimes\sigma_B}} \tr\ \sigma_B\ , \label{eq:minentropygeneraldefinition}
\end{align}
where we use the subscript $\rho$ to indicate what state the quantity refers to.
In~\cite{krs:entropy}, it was shown via semidefinite programming duality that
for a cq-state $\rho_{XQ}$, definition~\eqref{eq:minentropygeneraldefinition} of $\hmin(X|Q)$ coincides with definition~\eqref{eq:minentropyguess} in terms of the guessing-probability $P_{guess}(X|Q)$. The advantage of~\eqref{eq:minentropygeneraldefinition} is that it allows us to maximize over neighborhoods of $\rho_{XQ}$. This leads
to the definition of {\em smooth entropy}~\cite{renato:diss}, which is 
defined\footnote{Unlike in~\cite{renato:diss}, we require that half the $1$-norm is bounded. This ensures
that~$\hmin^\varepsilon(X|Q)_\rho \geq \hmin(X|Q)_\sigma$ if $\rho_{XQ} \approx_\eps \sigma_{XQ}$.}
as
\begin{align}
\hmineps(X|Q)_\rho :=\sup_{\substack{
\bar{\rho}_{XQ} \geq 0:\frac{1}{2}\|\bar{\rho}_{XQ}-\rho_{XQ}\|_1\leq \tr(\rho_{XQ})\cdot \varepsilon\\
\qquad\tr(\bar{\rho}_{XQ})\leq \tr(\rho_{XQ})
 }} \hmineps(X|Q)_{\bar{\rho}}\ .\label{eq:smoothminentropy}
\end{align}
We will also use the fact that if $\rho_{XQ}$ is a cq-state, the supremum can be restricted to density operators $\bar{\rho}_{XQ}$ where $X$ is classical and has the same range as the original~$X$. Definition~\eqref{eq:smoothminentropy} will be convenient for our proof: Roughly, we will construct some state that has
high min-entropy. We then show that the state created during a real execution of the protocol is $\eps$-close to this state.
By the above definition, the  actual state generated in the protocol has high \emph{smooth} min-entropy.

A useful property of the smooth min-entropy is that it obeys a chain rule~\cite[Theorem 3.2.12]{renato:diss}, which states
that
for any $ccq$-state $\rho_{XYQ}$, we have
\begin{align}\label{eq:chainrule}
\hmin^\eps (X|YQ)_\rho \geq \hmin^\eps (X|Q)_\rho -
\log |\cY|\;,
\end{align}
where $|\cY|$ is the size of the support of~$Y$.

\subsubsection{Uncertainty relations for post-measurement information}\label{sec:uncertainty}\label{sec:measurementpost}
When showing the security of weak string erasure, we need to consider a setting where an adversary can first extract some classical information~$K$
given access to a quantum system~$Q$ and later obtains some additional information~$\Theta$. His objective is to guess the value of a random variable~$X$.
Suppose he applies a measurement described by a POVM~$\{E_k\}_k$ to~$Q$, and retains only the measurement result $k$. We can think of this as a completely positive trace-preserving map (CPTPM) $\cK:\cB(\cH_Q)\rightarrow\cB(\cH_K)$. When he performs this measurement on the $Q$-part of a cq-state~$\rho_{XQ}$, we get
\begin{align*}
\rho_{X\cK(Q)}&:=(\id_X\otimes \cK)(\rho_{XQ})\\
&=\sum_k\tr_Q\left((\id_X\otimes E_k)\rho_{XQ}\right)\otimes\proj{k}\ ,
\end{align*}
which is a cc-state (i.e., an encoded joint distribution $P_{XK}$) if $X$~is classical. 
Due to its definition, the min-entropy $\hmin(X|Q)$ is intimately connected with such measurements, and in fact it is easy to see that
\begin{align}
\hmin(X|Q)=\min_{\cK} \hmin(X|\cK(Q))\ .\label{eq:measurementminimum}
\end{align}
This important identity relates min-entropies given {\em quantum} information~$Q$ to min-entropies given {\em classical} information $K=\cK(Q)$. 

Returning to the example given in~\eqref{eq:entropyuncertaintystate}, let us consider what happens if the adversary learns the basis information~$\Theta$
\emph{after} the measurement $\cK$. In~\cite[Theorem 4.7]{ww:pistar} it was shown that the minimal post-measurement min-entropy optimized over all 
measurements $\cK$ obeys
\begin{align*}
\min_\cK \hmin(X|\cK(Q)\Theta)&=-\log\left(\frac{1}{2}+\frac{1}{2\sqrt{2}}\right)\ ,
\end{align*}
which in the case of our example matches the min-entropy~$\hmin(X|Q)$ without post-measurement information~$\Theta$.
In our security proof, we will need to consider $n$~repetitions of the state~\eqref{eq:entropyuncertaintystate}, that is,
\begin{align*}
\rho_{X^n\Theta^nQ}&=\rho_{X\Theta Q}^{\otimes n}\  ,
\end{align*}
where $X^n=(X_1,\ldots,X_n)$ and $\Theta^n=(\Theta_1,\ldots, \Theta_n)$ are $n$-bit strings, and $\cH_Q\cong(\mathbb{C}^2)^{\otimes n}$. 
It follows from~\cite[Lemma 2]{noisy:arxiv} and~\cite{ww:pistar} that
\begin{align}
\min_\cK \hmin(X^n|\cK(Q)\Theta^n)&=-n\cdot \log\left(\frac{1}{2}+\frac{1}{2\sqrt{2}}\right)\ . \label{eq:postmeasurementminentropyuncertainty}
\end{align}
A generalization of this relation to smooth min-entropy is 
\begin{align}\label{eq:uncertaintyLargeN}
\begin{split}
\min_\cK \hmin^\eps(X^n|\cK(Q)\Theta^n) &\geq n \left(\frac{1}{2} - 2\delta\right)\\
\textrm{ where }\delta\in ]0,\frac{1}{2}[\ \textrm{ and }&
\eps = \exp\left(- \frac{\delta^2 n}{32(2 + \log\frac{1}{\delta})^2}\right)\ .
\end{split}
\end{align} 
This relation follows from~\cite[Corollary 3.4]{serge:new} using the standard purification trick (cf.~\cite[Lemma~2.3]{ww:compose}).
Our construction of a protocol for weak string erasure will make essential use of~\eqref{eq:postmeasurementminentropyuncertainty} and~\eqref{eq:uncertaintyLargeN}. 

\subsubsection{Secure keys and what it means to be ignorant}\label{sec:nonuniformity}
We will often informally say that an adversary ``does not know anything'' or ``does not learn anything'' or ``is ignorant'' about some random variable $X$, even when he holds some (quantum) information~$Q$.
In terms of the cq-state~$\rho_{XQ}$ this means that~$X$ is uniformly distributed on $\cX$, and independent of $Q$, that is,
\begin{align}\label{eq:cqstateindep}
\rho_{XQ}&=\tau_{\cX}\otimes\rho_Q\ .
\end{align} 
Clearly, for such a state, the uncertainty about $X$ given $Q$ is maximal, which in terms of the min-entropy means that
$\hmin(X|Q)=\log |\cX|$. 
For $\rho_{XQ}$ as in~\eqref{eq:cqstateindep}, 
$X$ is also referred to as an \emph{ideal key with respect to $Q$}.
 
In practice, we are generally forced to work with approximately ideal keys, where $X$~is called a  {\em $\varepsilon$-secure key with respect to $Q$} 
if $\rho_{XQ}$ is $\varepsilon$-close to the ideal state $\tau_{\cX} \otimes \rho_Q$, that is,
\begin{align}
\rho_{XQ} \approx_\eps \tau_{\cX} \otimes \rho_Q\ .\label{eq:limitinf}
\end{align}
This notion of a secure key behaves nicely under composition~\cite{BHLMO04,renato:compose,rk:locking}.

\subsection{Processes that increase uncertainty}
\label{sec:infoLoss}
To show the security of weak string erasure, we need to capture the amount of ``uncertainty'' that an adversary has as a result of his noisy storage $\cF$.
First, let us consider general processes which increase uncertainty. Note that from the definitions, it is
immediate~\cite[Theorem 3.1.12]{renato:diss}
that the min-entropy satisfies the following monotonicity property: for every CPTPM~$\cF:\cB(\cH_Q)\rightarrow\cB(\cH_{Q'})$, we have
\begin{align}
\hmin(X|\cF(Q))\geq \hmin(X|Q)\ .\label{eq:minentropymonotonicity}
\end{align}
An important case is where $\cH_Q=\cH_{Q_1Q_2}$ is bipartite, and $\cF=\tr_{Q_2}$ is the partial trace over the second system~$Q_2$. We then get
\begin{align}
\hmin(X|Q_1)\geq \hmin(X|Q_1Q_2)\ ,\label{eq:minentropymonotonicitypartialtrace}
\end{align}
reflecting the fact that ``forgetting'' information makes it harder to guess~$X$. 

Inequality~\eqref{eq:minentropymonotonicity} is insufficient for our purposes, and we will need a more quantitative estimate on the increase of entropy due to a channel~$\cF$ representing the adversary's memory. Clearly, such an estimate will depend on properties of~$\cF$. Here we express the bound in terms of the function~$P^\cF_{succ}(n)$ introduced in~\eqref{eq:succ}.
Intuitively, the following lemma shows that the uncertainty about~$X$ after application of $\cF$ to $Q$ is related to the problem of transmitting classical  bits
through the channel~$\cF$, where the number of bits is given by the min-entropy of~$X$.

\begin{lemma}\label{lem:minentropygeneration}
Consider an arbitrary cq-state $\rho_{XQ}$ and
a CPTPM $\cF:\cB(\cH_Q)\rightarrow\cB(\cH_{out})$. Then
 $\hmin(X|\cF(Q))\geq -\log P^{\cF}_{succ}(\lfloor H_\infty(X)\rfloor)$.
\end{lemma}
\begin{proof}
Let $k:=\lfloor H_\infty(X)\rfloor$. It is well-known (see e.g.,~\cite{Shaltiel}) that probability distributions $P_X$ with min-entropy at least~$k$ are convex combinations of ``flat'' distributions, i.e., uniform distributions over subsets of~$\cX$ of size $2^k$. In other words, there is a joint distribution  $P_{XT}$, where $T$ is distributed over subsets of size~$2^k$, such that
\begin{align*}
\begin{split}
P_X(x)&=\sum_{t} P_T(t)P_{X|T=t}(x)\ \textrm{ and }\\
P_{X|T=t}&\textrm{ is uniform on }t\subset \cX\ .
\end{split}
\end{align*}
The distribution $P_{XT}$ together with $\rho_{XQ}$ gives rise to a state $\rho_{XQT}$ whose partial trace is equal to $\rho_{XQ}$. Again using~\eqref{eq:minentropymonotonicitypartialtrace}, we get
\begin{align*}
\hmin(X|\cF(Q))\geq \hmin(X|\cF(Q)T)\ .
\end{align*}
By property~\eqref{eq:classicalconditioning} of the min-entropy when conditioning on classical information, we have
\begin{align}
\hmin(X|\cF(Q)T)=-\log \ExpE_{t\leftarrow P_T}\left[2^{-\hmin(X|\cF(Q),T=t)}\right]\ ,\label{eq:firstaux}
\end{align}
where $\ExpE_{t\leftarrow P_T}$ denotes the expectation value, and $\hmin(X|\cF(Q),T=t)$ is the min-entropy of the conditional state
\begin{align*}
\rho_{X\cF(Q)|T=t}&=\sum_{x} P_{X|T=t}(x)\proj{x}\otimes\cF(\rho_x)\ .
\end{align*}
Now we use the fact that $P_{X|T=t}$ is uniform over a set of size~$2^k$, and the definition of $P_{succ}^\cF(n)$. This leads to
\begin{align}
\begin{split}
\hmin(X|\cF(Q),T=t)\geq -\log P^{\cF}_{succ}(k)\\
\textrm{ for all }t\textrm{ in the support of }P_T\ .\end{split}\label{eq:secondaux}
\end{align}
Combining~\eqref{eq:firstaux} with~\eqref{eq:secondaux} gives the claim.
\end{proof}

We now give a straightforward but important generalization of this result.
\begin{lemma}\label{lem:basicminentropyincrease}
Consider an arbitrary ccq-state $\rho_{XTQ}$, and let $\varepsilon, \varepsilon'\geq 0$ be arbitrary. Let $\cF:\cB(\cH_Q)\rightarrow\cB(\cH_{Q_{out}})$  be an arbitrary CPTPM.  Then
\begin{align*}
\hmin^{\varepsilon+\varepsilon'}(X|T\cF(Q))\geq -\log P^{\cF}_{succ}\left(\big\lfloor \hmin^{\varepsilon}(X|T)-\log\frac{1}{\varepsilon'}\big\rfloor\right)\ .
\end{align*}
\end{lemma}
\begin{proof}
Clearly,
the statement for $\varepsilon=0$ implies the statement for any $\varepsilon>0$ because a CPTPM cannot increase distance. To prove the statement for $\varepsilon=0$, we consider the quantities~$2^{-\hmin(X|T=t)}$ of the conditional states $\rho_{X|T=t}$, together with the distribution~$P_T$ over~$\cT$ defined by the state~$\rho_{XTQ}$. Applying Markov's inequality $\Pr[Z\geq c]\leq \frac{\ExpE[Z]}{c}$ for any real-valued random variable~$Z$ and constant~$c>0$, we obtain
\begin{align*}
\begin{split}
\Pr_{t\leftarrow P_T}\left[2^{-\hmin(X|T=t)}\geq 2^{-\hmin(X|T)+\log\frac{1}{\varepsilon'}}\right]&\leq \\
\qquad \varepsilon'\left(2^{-\hmin(X|T)}\right)^{-1}\ExpE_{t\leftarrow P_T} \left[2^{-H_\infty(X|T=t)}\right]&=\varepsilon'\ .
\end{split}
\end{align*}
This implies that the distribution~$P_T$ has weight at least~$1-
\varepsilon'$ on the set \begin{align}
\good & = \left\{ t\in\cT\ |\
\hmin(X|T=t)\geq \big\lfloor \hmin(X|T)-\log\frac{1}{\varepsilon'}\big\rfloor\right\}\ .\label{eq:goodsetdef}
\end{align}
Accordingly, we can rewrite $\rho_{XTQ}$ as a convex combination 
\begin{align}
\begin{split}
\rho_{XTQ}&=(1-p) \cdot \rho_{XTQ|T\not\in\good}+p\cdot \rho_{XTQ|T\in\good}\ \textrm{ where }\\
p&=P_T(\good)\geq 1-\varepsilon'\ .\label{eq:convexcombrhogoodbad}
\end{split}
\end{align}
Set $\sigma_{XTQ}:=\rho_{XTQ|T\in\good}$. From~\eqref{eq:convexcombrhogoodbad}, we 
conclude that $\frac{1}{2}\|\rho_{XTQ}-\sigma_{XTQ}\|_1\leq 
\varepsilon'$. By the monotonicity of the distance under CPTPM, it therefore suffices to show that
\begin{align}
\hmin(X|T\cF(Q))_\sigma\geq -\log P^{\cF}_{succ}\left(\big\lfloor \hmin(X|T)_\rho-\log\frac{1}{\varepsilon'}\big\rfloor\right)\ .\label{eq:sigmaineqv}
\end{align}
For this purpose, note that $\sigma_{XT\cF(Q)}$ is given by the expression
\begin{align*}
\sigma_{TX\cF(Q)}&=\sum_{t\in\good} P_{T|T\in\good} (t) \proj{t}\otimes \rho_{X\cF(Q)|T=t}
\end{align*}
In particular, by using~\eqref{eq:classicalconditioning} again, we have
\begin{align}
2^{-\hmin(X|T\cF(Q))_\sigma}&=\ExpE_{t\leftarrow P_{T|T\in\good}}\left[2^{-\hmin(X|\cF(Q),T=t)_\rho}\right]\label{eq:sumhminxtf}
\end{align}
Using Lemma~\ref{lem:minentropygeneration} (applied to the conditional state $\rho_{XQ|T=t}$),  we conclude that
\begin{align}
\begin{split}
\hmin(X|\cF(Q),T=t)_\rho  \geq -\log P_{succ}^\cF\left(\lfloor \hmin(X|T)_\rho-\log\frac{1}{\varepsilon'}\rfloor\right)\\
\textrm{ for all }t\in\good\ .\hspace{20ex} &\label{eq:hminxfqtcond}
\end{split}
\end{align}
The claim~\eqref{eq:sigmaineqv} immediately follows from~\eqref{eq:hminxfqtcond} and~\eqref{eq:sumhminxtf}.
\end{proof}

\subsection{Defeating a quantum adversary: essential building blocks}

In order to build oblivious transfer and bit commitment from weak string erasure, we will employ three additional tools: first, we require {\em privacy 
amplification} against a quantum adversary~\cite{renato:diss,renato:compose} as explained in Section~\ref{sec:privacyamplification}. 
For oblivious transfer, we also need the notion of {\em min-entropy sampling} outlined in Section~\ref{sec:samplinghashing}. 
In particular, we discuss how min-entropy about classical information is approximately preserved when considering randomly chosen subsystems. We then show in
Section~\ref{sec:interactivehashing} how random subsets can be chosen in a cryptographically secure manner with a protocol called {\em interactive hashing}. 

\subsubsection{Privacy amplification}\label{sec:privacyamplification}
Intuitively, privacy amplification allows us to turn a long string $X$, about which the adversary holds some quantum information $Q$, into a shorter
string $Z = \Ext(X,R)$ about which he is almost entirely ignorant. The maximal length of this new string is directly related to the min-entropy $\hmin(X|Q)$ from 
Section~\ref{sec:advinformation}.
In order to obtain this new string, we will need a $2$-universal hash function: Formally, a function 
$\Ext:\sbin^n \otimes \cR \rightarrow\sbin^\ell $ is called $2$-universal if for all $x\neq x'\in\sbin^n$ and uniformly chosen $r \in_R \cR$, 
we have $\Pr[\Ext(x,r)=\Ext(x',r)]\leq 2^{-\ell}$.
\begin{theorem}[Privacy amplification~\cite{renato:diss,renato:compose}] \label{thm:PA} Consider a set of $2$-universal hash functions 
$\Ext: \sbin^n \otimes \cR \rightarrow \sbin^\ell$, and
a cq-state $\rho_{X^nQ}$, where $X^n$ is an $n$-bit string. Define $\rho_{X^n QR}=
\rho_{X^nQ}\otimes \tau_{\cR}$, i.e., $R$~is a random variable uniformly distributed on $\cR$, and independent of $X^n Q$. Then
\begin{align*}
\begin{split}
\rho_{\Ext(X^n,R)RQ} &\approx_{\eps'} \tau_{\sbin^\ell} \otimes \rho_{RQ} \\
\qquad\mbox{ where } 
\eps' &:=2^{-\frac{1}{2}\left(\hmineps(X^n|Q) - \ell \right)-1}+ 2\eps \
\end{split}
\end{align*}
for all $\varepsilon>0$.
\end{theorem}
It is important to stress that the extracted key~$\Ext(X^n,R)$ is secure even if the adversary is given~$R$ in addition to~$Q$. Theorem~\ref{thm:PA} immediately gives rise to a procedure allowing parties sharing some random variable~$X^n$ to extract a key secure against an adversary holding~$Q$. Indeed, one party can simply use independent randomness to pick~$r\in_R\cR$ uniformly at random, and distribute (publicly) the value of $r$. Because $2$-universal hash functions can be efficiently constructed (e.g., using linear 
functions~\cite{carterwegman77}), this {\em privacy amplification protocol} is efficient~\cite{BBR88,Impagliazzo:leftover,BBCM95}.

\subsubsection{Min-entropy sampling\label{sec:sampling}}\label{sec:minentropysampling}\label{sec:samplinghashing}
For oblivious transfer, we will make use of the sampling property of min-entropy which was first established by Vadhan~\cite{vadhan:sampling} in the classical case,
and in~\cite{kr:sampling} for the classical-quantum case.
Consider a cq-state $\rho_{X^nQ}$, where
$X^n=(X_1,\ldots X_n)$ is an $n$-bit string. An important property
of smooth min-entropy is that the {\em min-entropy rate}
\begin{align}
\frac{\hmin^\varepsilon(X^n|Q)}{n}\ \label{eq:minentropyrate}
\end{align}
is approximately preserved when considering a randomly chosen substring~$X_{\cS}$ of~$X^n$. In some sense, we can therefore think of~\eqref{eq:minentropyrate}
as the (average) min-entropy of an individual bit~$X_i$ given~$Q$. 

The corresponding technical statement is slightly more involved.
In essence, it requires to pick a subset~$\cS$ from a distribution~$P_{\cS}$ over subsets of~$[n]$ with certain properties ($P_{\cS}$ needs to be a so-called {\em averaging sampler},  see e.g.,~\cite{goldreich:samplers}). For concreteness, we consider the special case where $P_{\cS}$ is uniformly distributed over subsets of size $s=|\cS|$. Vadhan's result for the
classical case~\cite{vadhan:sampling} then shows that, for sufficiently large~$s$, we have
\begin{align*}
\frac{\hmineps(X_{\cS}|C)}{s}\geq \frac{\hmin(X^n|C)}{n}-\delta\ ,
\end{align*}
with high probability over the choice of~$\cS$, for some small $\varepsilon>0$ and $\delta>0$.  An analogous statement for the cq-case is given 
in~\cite{kr:sampling}. A major difference is that the result of~\cite{kr:sampling} for the quantum setting
requires~$X_i$ to be a block, i.e., a $\beta$-bit string instead of a single bit. 

Since our work is mainly a proof of principle, we do not yet care about optimality or efficiency. We therefore choose~$\cS$ to be uniform over all subsets of a fixed size~$s$. Furthermore, it is sufficient for our purposes to ensure that the min-entropy rate decreases by at most a factor of~$2$. 
Note that for technical reasons, the results in \cite{kr:sampling} requires the bit string to be partitioned into blocks of size $\beta$. A result in \cite{bitwiseSampling} shows however that the same bound must also hold for uniform bitwise sampling. This leads us to the following statement, which we derive by specializing the results of~\cite{kr:sampling} and combining it with the result in \cite{bitwiseSampling} (see appendix~\ref{sec:sampler} for details).

\begin{lemma}[Min-entropy sampling,~\cite{kr:sampling} combined with~\cite{bitwiseSampling}]\label{lem:samplingmodified} 
Let $\rho_{X^{m\beta} Q}$ be a cq-state, where $X^{m\beta}$ is an $m\beta$-bit string. 
Let
\begin{align*}
\frac{\hmin^{\varepsilon}(X^{m\beta}|Q)}{m\beta} \geq \lambda 
\end{align*}
be a lower bound on the smooth min-entropy rate of~$X^{m\beta}$ given~$Q$.
Let $\omega \geq 2$ be a constant, and assume $s,\beta\in\mathbb{N}$ are such that 
\begin{align}
s\geq m/4\qquad\textrm{and }\qquad \beta\geq \max \left
\{67,\frac{256 \omega^2}{\lambda^2} \right \}\ ,\label{eq:sbetacond}
\end{align}
and let
$P_\cS$ be the uniform distributions over subsets of~$[m \beta]$ of
size~$s \beta$.  Then
\begin{align*}
\begin{split}
\Pr_{\cS}\left[\frac{\hmin^{\varepsilon+4\delta}(X_{\cS}|Q)}{s\beta}\geq
\left(\frac{\omega-1}{\omega}\right)\lambda \right]&\geq 1-\delta^2\\
\textrm{ where }
\delta&=2^{-m \lambda^2/(512 \omega^2)}\ .
\end{split}
\end{align*}
\end{lemma}

\subsubsection{Aborting a protocol} \label{sec:abort}

As our protocols allow players to be malicious, they may abort simply by not sending a message. One way to handle this is to add a special symbol ``aborted'' to the definition of each primitive, and to handle this case separately in the protocol and the proof.
For simplicity, we will take a different approach here. Whenever a player does not send any message\footnote{Note that to decide whether Alice has sent a message requires to have an upper bound on the delivery time of a message.} (or a message that does not have the right format) the other player simply assumes that a particular valid message was sent, for example the string containing only zeros. Obviously, the malicious player could have sent this message himself, so refusing to send a message does not give any advantage to him. From now on we will therefore assume that all players always send a message when they are supposed to.

\subsubsection{Interactive hashing\label{sec:interactivehashing}}
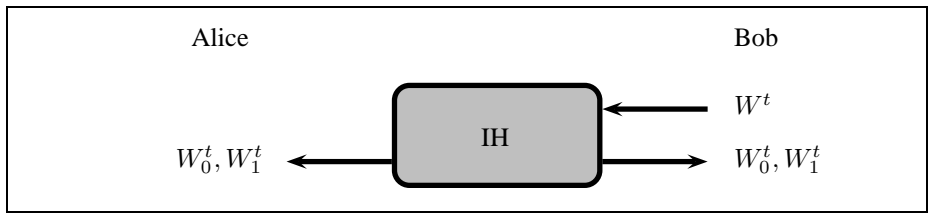
\begin{figure*}[t!]
\begin{center}
\scalebox{1.0}{
\fbox{\begin{pspicture}(-5,0)(7,2.5)
\psset{unit=.7cm}
\psset{linewidth=.8pt}
\psset{labelsep=2.5pt}
\put(0,0.3){\psframe[linewidth=2pt,framearc=.3,fillstyle=solid,
fillcolor=lightgray](4,2)}
\put(1.7,1.1){IH}
\put(4,1.8){\psline[linewidth=2pt]{<-}(2,0)}
\put(0,0.8){\psline[linewidth=2pt]{->}(-2,0)}
\put(4,0.8){\psline[linewidth=2pt]{->}(2,0)}
\put(-3.8,3){Alice}
\put(6.5,3){Bob}
\put(-4.1,0.7){$W_0^t,W_1^t$}
\put(6.5,0.7){$W_0^t,W_1^t$}
\put(6.5,1.7){$W^t$}
\end{pspicture}}
}
\end{center}
\caption{The concept of interactive hashing (IH): 
Honest Bob has input $W^t$. Interactive hashing creates substrings $W_0^t$ and $W_1^t$ such that
there exists $D \in \01$ with $W_D^t = W^t$, where $D$ is unknown to Alice, and Bob has little control over the choice of~$W_{1-D}^t$.}
\label{fig:ih}
\end{figure*}

A final tool we need is {\em interactive hashing}~\cite{ding,ding2,savvides:diss} first introduced in~\cite{IHfirst}. This is a two-party primitive where Bob inputs some  string $W^t$, and Alice has no input. The primitive then generates two strings $W_0^t$, $W_1^t$, 
with the property that one of the two equals~$W^t$. For a protocol implementing this primitive, security is intuitively specified by the following conditions: Alice does not learn which of the two strings
is indeed equal to~$W^t$. 
Conversely, Bob should have
very little control over the other string created by the protocol. Figure~\ref{fig:ih} depicts the idealized version of this primitive.

More formally, the following is achieved in~\cite[Theorem~5.6]{ding2}, where we refer 
to~\cite{savvides:diss} for the exact parameters used in the security
condition for Alice.
\begin{lemma}[Interactive Hashing~\cite{ding2,savvides:diss}] \label{lem:IH}
There exists a protocol called {\em interactive hashing~(IH)} between two
players, Alice and Bob, where Alice has no input, Bob has input
$W^t \in \sbin^t$ and both
players output $(W_0^t,W_1^t)\in\sbin^t\times\sbin^t$, satisfying the following:
\begin{enumerate}[]
\item{\em Correctness:} If both players are honest,
then $W_0^t\neq W_1^t$ and there exists a $D \in \{0,1\}$ such that
$W^t_D = W^t$.
Furthermore, the distribution of $W_{1-D}^t$ is uniform on 
$\sbin^t\backslash\{W^t\}$.
\item{\em Security for Bob:}
If Bob is honest, then $W_0^t\neq W_1^t$ and there exists a $D \in
\{0,1\}$ such that $W^t_D = W^t$. If Bob chooses $W^t$ uniformly at
random, then $D$ is uniform and independent of Alice's view.
\item{\em Security for Alice:}
If Alice is honest, then for every subset $\cS \subseteq \sbin^t$,
\begin{align*}
\Pr[W_0^t\in\cS\textrm{ and }W_1^t\in\cS]\leq 16 \cdot \frac{|\cS|}{2^t}
\end{align*}
\end{enumerate}
\end{lemma}

Note that even though this is not explicitly mentioned in~\cite{savvides:diss},
 aborts need to be treated as explained in Section~\ref{sec:abort}
to achieve Lemma~\ref{lem:IH}.

\section{Weak string erasure in the \nsw}\label{sec:wse}
We are now ready to introduce our main primitive. 
After giving a precise security definition in Section~\ref{sec:wseDef}, we present a protocol for realizing this primitive in 
the \nsw. We will subsequently show that the protocol satisfies the given security definition.

\subsection{Definition}\label{sec:wseDef}

\subsubsection*{``Strong'' versus weak string erasure}
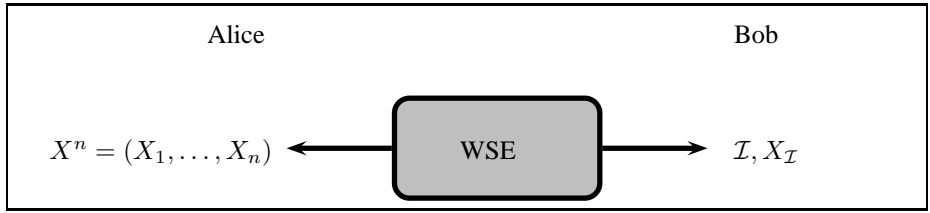
\begin{figure*}[t!]
\begin{center}
\scalebox{1.0}{
\fbox{\begin{pspicture}(-5,0)(7,2.5)
\psset{unit=.7cm}
\psset{linewidth=.8pt}
\psset{labelsep=2.5pt}
\put(0,0){\psframe[linewidth=2pt,framearc=.3,fillstyle=solid,
fillcolor=lightgray](4,2)}
\put(1.3,0.8){WSE}
\put(0,1){\psline[linewidth=2pt]{->}(-2,0)}
\put(4,1){\psline[linewidth=2pt]{->}(2,0)}
\put(-3.5,3){Alice}
\put(6.5,3){Bob}
\put(-6.5,0.8){$X^n=(X_1,\ldots,X_n)$}
\put(6.5,0.8){$\cI, X_{\cI}$}
\end{pspicture}} 
}
\end{center}
\caption{The ideal functionality of string erasure. The actual weak string erasure is somewhat weaker. However,  a 
dishonest party cannot gain significantly more
information from the protocol than provided by the ``box'' depicted above.}
\label{fig:wse}
\end{figure*}

In an ideal world, string erasure would realize the ideal functionality depicted in Figure~\ref{fig:wse}:
It takes no inputs, but provides Alice with a uniformly distributed string $n$-bit 
string $X^n=(X_1,\ldots,X_n)\in\sbin^n$, while Bob receives a random subset $\cI=\{i_1,\ldots,i_{|\cI|}\}\subset 2^{[n]}$ and the substring 
$X_{\cI}$.
The set of indices $\cI$ would be randomly distributed over all the set $2^{[n]}$ of all subsets 
of $[n]$. Intuitively, we think of the complement of $\cI$ as the locations of the ``erased'' bits.

Ideally, we would like to realize the functionality in Figure~\ref{fig:wse} in such a way that even a dishonest party cannot learn anything at all 
beyond what is provided by the box. Unfortunately, this definition is too stringent to be achieved by our protocol. We therefore relax our functionality to \emph{weak string erasure}, where the players may gain a small amount of additional information. 
More precisely, we allow a dishonest Bob to learn some information about~$X^n$ possibly different from~$(\cI,X_{\cI})$. However, we demand that his total information about $X^n$ is limited: given a dishonest Bob's system $B'$, he still has some residual amount of uncertainty about~$X^n$. 
For a dishonest Alice, we essentially retain the strong security property that she does not learn anything about the set of indices~$\cI$ that Bob receives. 
In order to obtain bit commitment and oblivious transfer later on, we also demand one additional property that may seem superfluous from a classical perspective, 
namely that Alice is ``committed'' to a choice of $X^n$ at the end of the protocol. This difficulty arises since unlike in a classical setting, a dishonest Alice
may for example store some quantum information and perform measurements only at a later time. This may allow her to determine parts of $X^n$ after the
protocol is completed. Security against such attacks is subtle to define in a quantum setting. To address this problem, we define security in terms of an 
``ideal'' state $\sigma_{A'X^n\cI X_{\cI}}$ that could have been obtained by an honest Alice by preparing some state on~$A'$ using~$X^n$ (i.e., by post-processing).  
Our security definition then demands that the actual state $\rho_{A'B}$ shared by dishonest Alice and honest Bob after the execution of the protocol 
has the same form as the partial trace of the ideal state, that is, $\rho_{A'B} = \sigma_{A'\cI X_{\cI}}$.

\subsubsection*{Formal definition}
In the following definition of weak string erasure, we write $\rho_{AB}$ for the resulting state at the end of the protocol if both parties are honest, $\rho_{A'B}$ is Alice is dishonest and $\rho_{AB'}$ if Bob is dishonest. Our definition is phrased in terms of ideal states denoted by $\sigma$ that exhibit
all the desired properties of weak string erasure. We then demand that the actual states $\rho$ created during a real execution of the protocol
are at least $\eps$-close to such ideal states no matter what kind of attack the dishonest party may perform.

\begin{definition}\label{def:wse}
An {\em $(n,\lambda,\varepsilon)$-weak string erasure (WSE) scheme}  is a protocol between Alice and Bob satisfying the following properties:
\begin{enumerate}[]
\item{\em Correctness:}
If both parties are honest, then the ideal state $\sigma_{X^n \cI X_{\cI}}$ is defined such that
\begin{enumerate}
\item The joint distribution of the $n$-bit string $X^n$ and the subset $\cI$ is uniform:
\begin{align}
\sigma_{X^n\cI}&=\tau_{\sbin^n}\otimes\tau_{2^{[n]}}\ ,
\end{align}
\item The joint state $\rho_{AB}$ created by the real protocol is equal to the ideal state:
\begin{align}
\rho_{AB}=\sigma_{X^n\cI X_\cI}\ .\label{eq:strongerasure}
\end{align}
where we identify $(A,B)$ with $(X^n,\cI X_{\cI})$.
\end{enumerate}
\item{\em Security for Alice:} If Alice is honest, then there exists an ideal state
$\sigma_{X^nB'}$ such that
\begin{enumerate}
\item The amount of information $B'$ gives Bob about $X^n$ is limited:
\begin{align*}
\frac{1}{n}\hmin(X^n|B')_\sigma\geq \lambda
\end{align*}
\item
The joint state $\rho_{AB'}$ created by the real protocol is $\eps$-close to the ideal state:
\begin{align*}
\sigma_{X^nB'} \approx_{\eps} \rho_{AB'}
\end{align*}
where we identify $(X^n,B')$ with $(A,B')$.
\end{enumerate}

\item{\em Security for Bob:} If Bob is honest, 
then there exists an ideal state $\sigma_{A'\widehat{X}^n\cI}$, where $\widehat{X}^n \in \01^n$ and $\cI \subseteq [n]$ such that
\begin{enumerate}
\item The random variable $\cI$ is independent of $A'\widehat{X}^n$ and uniformly distributed over $2^{[n]}$:
\begin{align*}
\sigma_{A'\widehat{X}^n\cI}=\sigma_{A'\widehat{X}^n}\otimes\tau_{2^{[n]}}\ .
\end{align*} 
\item
The joint state $\rho_{A'B}$ created by the real protocol condition on the event that Alice does not abort is equal to the ideal state:
\begin{align*}
\rho_{A'B}= \sigma_{A'(\cI \widehat{X}_\cI)}
\end{align*}
where we identify $(A',B)$ with $(A',\cI\widehat{X}_\cI)$.
\end{enumerate}
\end{enumerate}
\end{definition}
Note that we do not require $X^n$ to be uniform when Bob is dishonest.
To show security of bit commitment and oblivious transfer we will only require that $X^n$ has high min-entropy.
The condition that the real state is close
to an ideal state having high min-entropy means that the real state has smooth min-entropy as outlined in Section~\ref{sec:prelim}.

\subsection{Protocol}
\begin{figure*}[t!]
\begin{center}
\qquad\Qcircuit 
@C=1em @R=1.0em { 
&&&&\lstick{x_1,\ldots,x_n} &\qw&\gate{H} &\qw&\qw&\qw&\qw &\qw&\qw& \qw&\qw&\qw &\gate{H} &\qw&\qw&\meter&\qw&\qw&\qw&\ustick{\tilde{x}_1,\ldots,\tilde{x}_n}\qw&\qw&\qw&\qw&\qw& \qw&\qw&\qw&\gate{\bot} & \qw\\
&&&&&&&&&&&&&&\lstick{\tilde{\theta}_1, \ldots,\tilde{\theta}_n}&\qw&\ctrl{-1}&\qw&\qw&\qw&\qw&\qw&\qw&\qw &\qw&\qw&\qw&\qw&\qw&\gate{\oplus}&\qw&\ctrl{-1}&\qw\\
&&&&&&&&&&&&&&&&&\mbox{Honest Bob (time $t$)}&&&&&&&&&&&&&&&\\
&&&&\lstick{\theta_1 ,\ldots ,\theta_n}&\qw&\ctrl{-3}&\qw&\qw&\qw &\qw&\qw &\qw&\qw &\qw&\qw &\qw&\qw&\qw&\qw&\qw &\qw&\qw&\qw & \qw& \qw&\qw&\qw&\qw&\ctrl{-2}&\qw&\qw&\qw\\
%
&&&&\mbox{Honest Alice}&&&&&&&&&&& & && &&&&&&&&&&&& \mbox{Honest Bob}&&\\
&&&&&& &&&&&&&& & &&&& &&   &&&&&&&&&\mbox{(time $t+\Delta t$)}\gategroup{1}{1}{4}{7}{5em}{.}\gategroup{1}{12}{2}{25}{4.5em}{.}\gategroup{1}{30}{4}{32}{5em}{.} &&&
} 
\end{center}
\caption{The protocol as a circuit. Alice chooses a random string $x^n = (x_1,\ldots,x_n) \in \01^n$. She then encodes the bits in random bases specified by~$\theta^n = (\theta_1,\ldots,\theta_n) \in \01^n$
and sends the corresponding quantum states to Bob. Bob measures in random bases specified by~$\tilde{\theta}^n = (\tilde{\theta}_1,\ldots,\tilde{\theta}_n) \in \01^n$ obtaining measurement outcomes 
$\tilde{x}^n = (\tilde{x}_1,\ldots,\tilde{x}_n)$. 
Upon reception of the basis string $\theta^n$, Bob determines the locations where he measured in the same basis by computing
the bit-wise xor $\theta^n \oplus \tilde{\theta}^n=(\theta_1\oplus\tilde{\theta}_1,\ldots,\theta_n\oplus\tilde{\theta}_n)$. He subsequently discards the bits he measured in the wrong bases (indicated by $\bot$: this replaces the classical input symbol by an erasure symbol).\label{fig:protocolhonestparties}}
\end{figure*}
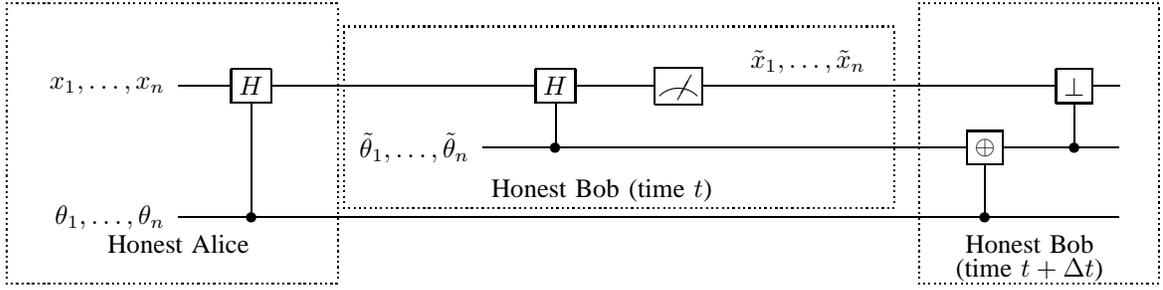

We now consider a simple protocol achieving weak string erasure in the \nsw\ using BB84-states.
Other encodings are certainly possible, and we will discuss some of the implications of 
this choice of encoding in Section~\ref{sec:extensions}. 
This protocol is essentially identical to the first step of known protocols 
for quantum key distribution~\cite{wiesner:conjugate,bb84}. However, as explained in the last section, our security requirements differ greatly as
we are dealing with two mutually distrustful parties.

\begin{protocol}{1}{Weak String Erasure (WSE)}{Outputs: $x^n \in \sbin^n$ to Alice, $(\cI,z^{|\cI|}) \in 2^{[n]} \times \sbin^{|\cI|}$ 
to Bob.}\label{proto:wse}
\item[1.] {\bf Alice:} Chooses a string $x^n \in_R \01^n$ and bases-specifying string $\theta^n \in_R \01^n$ uniformly at random. She encodes each bit $x_i$ in 
the basis given by $\theta_i$ (i.e., as $H^{\theta_i}\ket{x_i}$) and sends it to Bob.

\item[2.] {\bf Bob: } Chooses a basis string $\tilde{\theta}^n \in_R \01^n$ uniformly at random. When receiving the $i$-th qubit, Bob measures it in the basis
given by $\tilde{\theta}_i$ to obtain outcome $\tilde{x}_i$.

\item[Both parties wait time $\Delta t$.]

\item[3.] {\bf Alice: } Sends the basis information $\theta^n$ to Bob, and outputs $x^n$.
\item[4.] {\bf Bob: } Computes $\iSet \assign \{i \in [n] \mid \theta_i = \tilde{\theta}_i\}$, and outputs $(\cI,z^{|\cI|}):=(\cI,\tilde{x}_{\cI})$.
\end{protocol}

Our main claim is the following:
\begin{theorem}[Weak string erasure]\label{thm:mainclaimwse}
\begin{enumerate}[(i)]
\item
Let $\delta\in ]0,\frac{1}{2}[$ and 
let Bob's storage be given by $\cF: \bop(\hin) \rightarrow \bop(\hout)$.
Then Protocol~1 is an $(n,\lambda(\delta,n),\eps(\delta,n))$-weak string erasure protocol with
min-entropy rate
\begin{align*}
\lambda(\delta,n)=-\frac{1}{n} \log P^\cF_{succ}\left(\left(\frac{1}{2}-\delta\right)\cdot n\right)\ ,
\end{align*}
and error
\begin{align}
\eps(\delta,n) =2\exp \left(-\frac{\delta^2}{512(4+\log\frac{1}{\delta})^2}\cdot  n\right)\ .\label{eq:thmepsdef}
\end{align}
\item
Suppose $\cF=\cN^{\otimes \nu n}$ for a storage rate~$\nu>0$, $\cN$ satisfying the \scp~\eqref{eq:strongconverseproperty} and having capacity~$C_\cN$ bounded by
\begin{align*}
C_\cN\cdot\nu<\frac{1}{2}\ .
\end{align*} 
Let $\delta\in ]0,\frac{1}{2}-C_\cN\cdot \nu[$.
Then Protocol~1 is an $(n,\tilde{\lambda}(\delta),\eps(\delta,n))$-weak string erasure protocol 
for sufficiently large~$n$, where 
\begin{align*}
\tilde{\lambda}(\delta) &=\nu \cdot \gamma^\cN\left(\frac{\textfrac{1}{2}-\delta}{\nu}\right)\ .
\end{align*}
\end{enumerate}
\end{theorem}

It is easy to see that that the protocol is correct if both parties are honest:
if Alice is honest, her string $X^n=x^n$ is chosen uniformly at random from $\01^n$ as desired, and if Bob is honest, he will clearly
obtain $\tilde{x}_i = x_i$ whenever $i \in \iSet$ for a random subset $\iSet \subseteq [n]$. 
The remainder of Section~\ref{sec:wse} is thus devoted to proving security if one of the parties is dishonest: 
In Section~\ref{sec:alicesecurity}, we use the properties of the channel~$\cF$ to show that the protocol is secure against a dishonest 
Bob. In Section~\ref{sec:bobsecurity}, we argue that the protocol satisfies Definition~\ref{def:wse} when Alice is dishonest.

\subsection{Security for honest Alice~\label{sec:alicesecurity}} 

We now show that for any cheating strategy of a dishonest Bob, his min-entropy about the string $X^n =(X_1,\ldots,X_n)$ is large.
Before turning to the proof, we first explain in Figure~\ref{fig:bobattackfigure} how our model restricts the actions of Bob in our protocol. 
At time~$t$, Bob receives an encoding of a classical string $x^n = (x_1,\ldots,x_n)$ which he would like to reconstruct as accurately as possible.
To this end, he can apply any CPTPM $\cE:\cB((\mathbb{C}^2)^{\otimes n})\rightarrow\cB(\cH_{in}\otimes\cH_{K})$ with the following property: For any input state $\rho$ on 
$(\mathbb{C}^2)^{\otimes n}$, he obtains an output state $\zeta_{Q_{in}K}=\cE(\rho)$, where $Q_{in}$ is the quantum information he will put into his quantum
storage, and~$K$ is any additional classical information he retains. Note that we allow an arbitrary amount of classical storage, that is, 
$\cH_K$ may be arbitrarily large\footnote{It is sufficient for any adversary to store~$2^n$ bits, one for each possible
basis string $\Theta^n$~\cite{ww:pistar}.}. We call the map~$\cE$ Bob's {\em encoding attack}.

We can think of the encoding attack~$\cE$ as being composed of two steps, $\cE=(\id_{Q_{in}}\otimes \cK)\circ \cE_{1}$ where Bob first applies an arbitrary CPTPM~$\cE_{1}:\cB((\mathbb{C}^2)^{\otimes n})\rightarrow \cB(\cH_{Q_{in}}\otimes\cH_{\tilde{Q}})$, and subsequently performs a 
measurement $\cK:\cB(\cH_{\tilde{Q}})\rightarrow\cB(\cH_{K})$ on $\cH_{\tilde{Q}}$. The outcome of this measurement forms his classical information $K = \cK(\tilde{Q})$. 
For example, Bob can measure some of the incoming qubits, or encode some information using an error-correcting code.
The joint state before his storage noise is applied is
hence given by
\begin{align}
\begin{split}
\rho_{X^n\Theta^n K Q_{in}}=\hspace{42ex} &\label{eq:bobstorageattack}\\
\frac{1}{(2^n)^2}\hspace{-5ex}\sum_{\substack{\qquad x^n,\theta^n\in\sbin^n\\ k \in \cK}} 
\hspace{-3ex}P_{K|X^n=x^n,\Theta^n=\theta^n}(k)
\underbrace{\pi_{x^n}\otimes\pi_{\theta^n}}_{\rm Alice}
\otimes \underbrace{\pi_{k}\otimes \zeta_{x^n\theta^n k}}_{\rm Bob}\ 
\end{split}
\end{align}
where $\zeta_{x^n \theta^n k}$ is the conditional state on $\cH_{in}$ conditioned on the string $X^n=x^n$, the basis choice
$\Theta^n=\theta^n$ and Bob's classical measurement outcome $K=k$. Here we used the abbreviation $\pi_x:=\proj{x}$. The state~\eqref{eq:bobstorageattack} is completely determined by Bob's encoding attack~$\cE$ at time~$t$.  

Bob's storage~$Q_{in}$ then undergoes noise described by~$\cF:\cB(\cH_{in})\rightarrow\cB(\cH_{out})$, and the state evolves to $\rho_{X^n \Theta^n K \cF(Q_{in})}$. 
At time $t+\Delta t$, Bob additionally receives the basis information $\Theta^n = \theta^n$. The joint state is now given by
\begin{align}
\begin{split}
\rho_{X^n\Theta^n K \cF(Q_{in})}=\hspace{42ex} &\\
\hspace{-1ex}\frac{1}{(2^n)^2}\hspace{-7ex}\sum_{\substack{\ \qquad x^n,\ \theta^n\in\sbin^n\\ k \in \cK}}
\hspace{-3ex}P_{K|X^n=x^n,\Theta^n=\theta^n}(k)
\underbrace{\pi_{x^n}}_{\rm Alice}\otimes\underbrace{\pi_{\theta^n}\otimes\pi_{k}\otimes \cF(\zeta_{x^n\theta^n k})}_{{\rm Bob\ } B'}\ ,\label{eq:statetoanalyze}
\end{split}
\end{align}
where Bob holds $B' = \Theta^n K \cF(Q_{in})$.
We now show that Bob's information~$B'$ about~$X^n$ is limited for large~$n$.

\begin{center}  
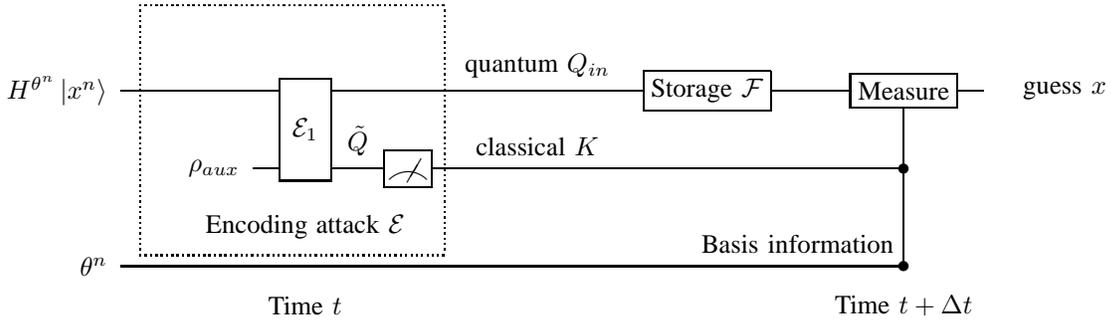
\begin{figure*}[t!] 
\qquad\hspace{2cm}
\Qcircuit @C=1em @R=1.5em {
\lstick{H^{\theta^n}\ket{x^n}}
\gategroup{1}{5}{2}{8}{6.5em}{.}
&\qw&\qw&\qw&\qw&\qw&\multigate{1}{\mbox{$\cE_1$}}&\qw&\qw&\qw&\qw&\qw&\ustick{\mbox{quantum $Q_{in}$}}\qw&
\qw&\qw&\qw&\gate{\mbox{Storage $\cF$}}&\qw&\qw&\gate{\mbox{Measure}}&\qw&\rstick{\mbox{guess $x$}}\\  
&&&&&\lstick{\rho_{aux}}&\ghost{\cE_1} &\ustick{\tilde{Q}}\qw&\meter&\qw&\qw&\qw&\ustick{\mbox{classical $K$}}\qw&\qw&\qw&\qw&\qw&\qw&\qw
&\ctrl{-1}&&\\
&&&&&&\mbox{Encoding attack $\cE$}&&&&&&&&&&&
&&\\
\lstick{\theta^n} &\qw&\qw&\qw&\qw&\qw&\qw&\qw&\qw&\qw&\qw&\qw&\qw&\qw&\qw&\qw&\qw&\ustick{\mbox{Basis information}}\qw&\qw&
\ctrl{-3}&&\\
&&&&&&\mbox{Time $t$}&&&&&&&&&&&&&\mbox{Time $t + \Delta t$}
}\vspace{2ex}
\caption{The most general structure of a cheating Bob. Bob's action at time~$t$ consists of a CPTPM~$\cE_1$, followed by a (partial) measurement $\cK$, where
he may use an arbitrary ancilla $\rho_{aux}$. 
At time $t + \Delta t$, Bob can try to reconstruct $x^n = (x_1,\ldots,x_n)$ given 
the content $\cF(Q_{in})$ of the storage device,
the classical measurement result $K=\cK(\tilde{Q})$, and the basis information $\theta^n = (\theta_1,\ldots,\theta_n)$.\label{fig:bobattackfigure}}
\end{figure*}
\end{center}

\begin{theorem}[Security for Alice]\label{thm:WSEdishonestBob}
Fix $\delta\in ]0,\frac{1}{2}[$ and let 
\begin{align*}
\eps=2\exp \left(-\frac{(\delta/4)^2}{32(2+\log\frac{4}{\delta})^2}\cdot  n\right)\ .
\end{align*}
Then for any attack of a dishonest Bob with storage~$\cF:\cB(\cH_{in})\rightarrow\cB(\cH_{out})$,
there exists a cq-state $\sigma_{X^nB'}$ such that
\begin{enumerate}
\item $\sigma_{X^n B'} \approx_{\eps} \rho_{X^n B'}$,\\2mm
\item $\frac{1}{n}\hmin(X^n|B')_{\sigma}\geq -\frac{1}{n}\log
P_{succ}^{\cF}\left(\left(\frac{1}{2} - \delta\right)n\right)$,\\2mm
\end{enumerate}
where $\rho_{X^nB'}$ is given by~\eqref{eq:statetoanalyze}.
In particular, if, for some $R<\frac12$, we have
$\lim_{n\rightarrow\infty}-\frac{1}{n}\log P^{\cF}_{succ}(nR)>0$, then
$\rho_{X^nB'}$ is exponentially close (in $n$)
to a state~$\sigma_{X^nB'}$ with constant min-entropy rate
$\frac{1}{n}\hmin(X^n|B')_\sigma$.
\end{theorem}
\begin{proof}
We use the notation introduced in~\eqref{eq:statetoanalyze}. By definition~\eqref{eq:smoothminentropy} of the smooth min-entropy, statements~(1) and~(2)  follow if we show that  the smooth min-entropy rate~$\frac{1}{n}\hmin^\varepsilon(X^n|B')_\rho$ is lower bounded by the expression on the rhs.~in~(2).
By the uncertainty relation~\eqref{eq:uncertaintyLargeN}, we have
\begin{align*}
\hmin^{\eps/2}(X^n|\Theta^n K)_\rho\geq \frac{n}{2}-\frac{n \delta}{2}\ .
\end{align*}
Using Lemma~\ref{lem:basicminentropyincrease} applied to $T=(\Theta^n,K)$, we conclude that 
for $Q_{out}=\cF(Q_{in})$ after the noise~$\cF$ there exists the claimed ideal state and 
\begin{align*}
\hmin^{\eps}(X^n|\Theta^n KQ_{out})_\rho&\geq -\log P_{succ}^\cF
\left(\frac{n}{2} - \frac{n \delta}{2}-\log\frac{2}{\eps}\right)\\
& \geq - \log P_{succ}^\cF\left(\frac{n}{2} - \frac{n\delta}{2} - \frac{n \delta}{2}\right)\ ,
\end{align*}
where 
the final inequality follows from the monotonicity property of the success probability
$P^{\cF}_{succ}(m)\leq P^{\cF}_{succ}(m')$ for $m\geq m'$ and the fact that
 $\log \frac{2}{\varepsilon}\leq \frac{\delta}{2} n$ because $(\delta/4)^2/(32(2-\log\delta/4)^2) \leq \delta/2$ for any
$0 < \delta < 1/2$.
\end{proof}

Let us specialize Theorem~\ref{thm:WSEdishonestBob} to the case where $\cF$ is a tensor product channel.
\begin{corollary}\label{cor:WSEdishonestBob}\label{cor:tensorChannel}
Let Bob's storage be described by $\cF = \cN^{\otimes \nu n}$ with $\nu > 0$, where $\cN$ satisfies the \scp~\eqref{eq:strongconverseproperty},
and 
\begin{align*}
C_\cN\cdot \nu< \frac{1}{2}\ .
\end{align*}
Fix $\delta\in ]0,\frac{1}{2}-C_\cN\cdot \nu[$, and let $\eps=\eps(\delta,n)$ be defined 
by~\eqref{eq:thmepsdef}. Then for any attack of a dishonest Bob, there exists a cq-state $\sigma_{X^nB'}$ such that
\begin{enumerate}
\item $\sigma_{X^n B'} \approx_{\eps} \rho_{X^n B'}$,
\item $\frac{1}{n}\hmin(X^n|B')_{\sigma}\geq \nu\cdot \gamma^\cN\left(\frac{1/2-\delta}{\nu}\right)>0$,
\end{enumerate}
where $\rho_{X^nB'}$ is given by~\eqref{eq:statetoanalyze}.
\end{corollary}
\begin{proof}
Substituting $n$ by $\nu n$ and $R$ by $R/\nu$, the \scp~\eqref{eq:strongconverseproperty} turns into
\begin{align*}
-\frac{1}{n}\log P^{\cN^{\otimes \nu n}}_{succ}(nR)\geq \nu \cdot \gamma^{\cN}(R/\nu)\ 
\end{align*}
for sufficiently large~$n$.
The claim then follows from Theorem~\ref{thm:WSEdishonestBob} by setting~$R:=\frac{1}{2}-\delta$.
\end{proof} 
Theorem~\ref{thm:WSEdishonestBob} and Corollary~\ref{cor:WSEdishonestBob} 
establish the first part of Theorem~\ref{thm:mainclaimwse}. It remains to analyze the security against a dishonest Alice. 
 
\subsection{Security for honest Bob\label{sec:bobsecurity}}
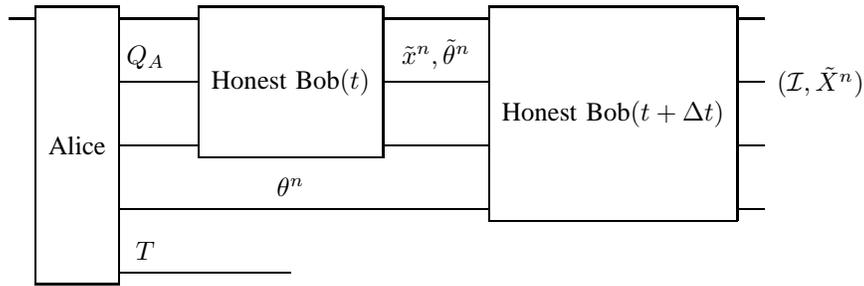
\begin{figure*}[t!]
\begin{center}
\hspace{1.5cm}\Qcircuit @C=1em @R=1.5em { 
&\multigate{4}{\mbox{Alice}}\pureghost{}&\qw &\qw          &\multigate{2}{\mbox{Honest Bob}(t)}&\qw&\qw  &\qw&\multigate{3}{\mbox{Honest Bob}(t+\Delta t)}  &\qw\\
&\pureghost{\mbox{Alice}}&\ustick{Q_A}\qw   &\qw          &\ghost{\mbox{Honest Bob}(t)}&\qw    &\ustick{\tilde{x}^n,\tilde{\theta}^n}\qw  &\qw &\ghost{\mbox{Honest Bob}(t+\Delta t)}&\rstick{(\cI,\tilde{X}^n)}\qw\\
&\pureghost{\mbox{Alice}}&\qw                &\qw          &\ghost{\mbox{Honest Bob}(t)}&\qw   &\qw   &\qw&\ghost{\mbox{Honest Bob}(t+\Delta t)}&\qw\\
&\pureghost{\mbox{Alice}}&\qw&\qw&\ustick{\theta^n}\qw    &\qw&\qw     &\qw &\ghost{\mbox{Honest Bob}(t+\Delta t)}&\qw\\ 
&\pureghost{\mbox{Alice}}&\ustick{T}\qw     &\qw    &\qw      \\
}
\end{center} 
\caption{This circuit shows the interaction between a dishonest party Alice and an honest Bob: Alice sends some $n$-qubit register~$Q_A$ and $n$~classical bits 
$\theta^n$ to Bob, and also retains some possibly quantum register~$T$. Honest Bob computes $\cI$ and $\tilde{X}^n$ as before. 
This generates an overall state $\rho_{\Theta^n T\cI \tilde{X}^n}$, where Alice's information~$A'$ after execution consists of~the 
classical string~$\Theta^n$ and~$T$. \label{fig:dishonestaliceinteraction}} 
\end{figure*}

When Alice is dishonest, it is intuitively obvious that she is unable to gain any information about the index set $\cI$, since she never receives 
any information from Bob during our protocol. Yet, in order to obtain bit commitment and oblivious transfer from weak string erasure we 
require a more careful security analysis. Figure~\ref{fig:dishonestaliceinteraction} depicts the form of any interaction between a cheating Alice and an honest
Bob. Since Alice takes no input in the protocol, her actions are completely specified by the state~$\rho_{Q_A\Theta^n T}$ she outputs, 
where $\cH_{Q_A}\cong (\mathbb{C}^2)^{\otimes n}$ is an $n$-qubit register that she sends to Bob (in the case where Alice is honest, this encodes the string $X^n$), $\Theta^n$ is some classical $n$-bit string (in the case where Alice is honest, this encodes the bases), and $\cH_{T}$ is an auxiliary register of Alice corresponding to the (quantum) information she holds after execution of the protocol. In the actual protocol, an honest Bob proceeds as shown 
in Figure~\ref{fig:protocolhonestparties}, that is, 
\begin{enumerate}
\item
Upon receipt of $Q_A$ at time~$t$, an honest Bob measures in randomly chosen bases specified by the string $\tilde{\Theta}^n=(\tilde{\Theta}_1,\ldots,\tilde{\Theta}_n)\in\sbin^n$, 
obtaining measurement outcomes $\tilde{X}^n=(\tilde{X}_1,\ldots, \tilde{X}_n)$.
\item After receiving $\Theta^n=(\Theta_1,\ldots, \Theta_n)$ at time $t+\Delta t$, he computes the intersecting set $\cI$ defined by $\tilde{\Theta}^n$ and~$\Theta^n$, and the corresponding substring $\tilde{X}_\cI$. 
\end{enumerate}
The protocol thus creates some state $\rho_{A'\cI \tilde{X}_{\cI}}$, where $A'=(\Theta^n T)$ is Alice's information, and $B=(\cI \tilde{X}_{\cI})$ is the 
information obtained by Bob.  Note that this state can be obtained from $\rho_{A'X^n\Theta^n\tilde{\Theta}^n}$ because $\cI$ is a function of~$\Theta^n$ 
and $\tilde{\Theta}^n$, and $\tilde{X}_\cI$ is a function of $\tilde{X}^n$ and $\cI$. 

\begin{theorem}
[Security for Bob] Protocol~1 satisfies security for honest Bob.
\end{theorem}

\begin{proof}
We now construct a state $\sigma_{A'\hat{X}^n\cI}$ with the required properties. 
For simplicity, we give an algorithmic description of this state. It is obtained by letting Alice and Bob interact with a simulator which has perfect
quantum memory. 
Note that this simulator
is purely imaginary and is merely used to specify the desired ideal state $\sigma_{A'\hat{X}^n \cI}$. However, we will
later show that the real state created during the protocol equals this ideal state on the registers held by Alice and Bob. 
Figure~\ref{fig:dishonestaliceinteractionsim} summarizes the actions of the simulator:

\begin{figure*}[t!]
\begin{center}
\qquad\Qcircuit @C=1em @R=1.5em { 
&\multigate{2}{\mbox{Alice}}&\qw&\ustick{Q_A}\qw&\qw & \qw &\qw&\gate{H}&\qw&\qw&\meter & \qw &\qw
&\ustick{\hat{x}^n}\qw& \qw&\qw     &\gate{H}&\qw&
\ustick{\hat{Q}_A}\qw
&\qw  &\gate{\mbox{Bob}(t)}   &\qw    &\ustick{\tilde{x}^n,\tilde{\theta}^n}\qw  &\qw &\multigate{1}{\mbox{Bob}(t+\Delta t)}&\qw\\
&\pureghost{\mbox{Alice}}&\qw&\qw&\ustick{\theta^n}\qw&\qw&\qw& \ctrl{-1}& \qw&&   & & &  
&\lstick{\hat{\theta}^n}&\qw&\ctrl{-1}     &\qw     & \qw&\qw &\qw&\qw&\qw&\qw&\ghost{\mbox{Bob}(t+\Delta t)}\\
&\pureghost{\mbox{Alice}}&\qw&\ustick{T}\qw&\qw & & &    &     & &   & &    &\mbox{Simulator}\gategroup{1}{8}{3}{19}{2em}{.}        &  & & & & & \\
}
\end{center}  
\caption{In the security proof, we put an intermediate ``simulator'' between Alice and Bob to generate the state $\sigma_{A'\hat{X}^n\cI}$. We will show the security definition~\ref{def:wse} is satisfied with~$\sigma_{A'\hat{X}^n\cI}$. The simulator measures the quantum register in the basis specified by the bit string. He then encodes the measurement result $\hat{X}^n = (\hat{X}_1,\ldots, \hat{X}_n)$ 
into randomly chosen bases. \label{fig:dishonestaliceinteractionsim}}
\end{figure*}
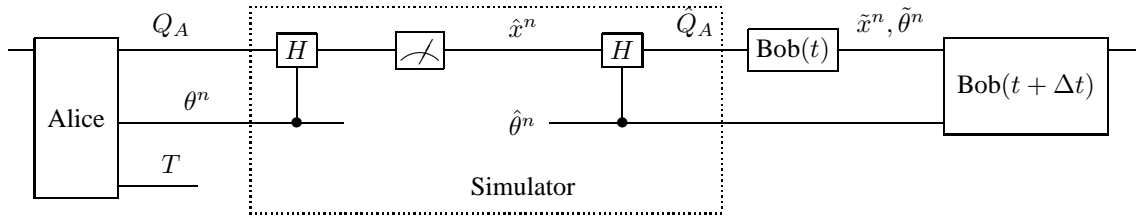

\begin{enumerate}
\item
First, the simulator measures the $n$-qubits $Q_A$ in the bases specified by the bits $\Theta^n=(\Theta_1,\ldots,\Theta_n)$, obtaining 
measurement outcomes $\hat{X}^n=(\hat{X}_1,\ldots,\hat{X}_n)$. 
\item
Second, the simulator re-encodes the measurement outcomes $\hat{X}^n$ using randomly chosen bases specified by $\hat{\Theta}^n=(\hat{\Theta}_1,\ldots,\hat{\Theta}_n)\in_R\sbin^n$. 
He then sends the corresponding qubits to Bob (i.e., the states $H^{\hat{\theta}_i}\ket{\hat{x}_i}$). We call this quantum register $\hat{Q}_A$. 
\item
Finally, the simulator provides Bob with the basis string $\hat{\Theta}^n=(\hat{\Theta}_1,\ldots,\hat{\Theta}_n)$. 
\end{enumerate}
An honest Bob proceeds as before, but with $\Theta^n$ replaced by the simulator's string~$\hat{\Theta}^n$, and $Q_A$ replaced by the  simulator's
quantum message $\hat{Q}_A$. 
As before, Alice's 
information~$A'=(T\Theta^n)$ consists of the string~$\Theta^n$ and her (quantum) system~$T$. 
The state $\sigma_{A' \cI \tilde{X}_{\cI}}$ held by Alice and Bob can be obtained 
from~$\sigma_{A'\hat{X}^n\hat{\Theta}^n\tilde{\Theta}^n}$, noting
that $\hat{X}_{\cI} = \tilde{X}_{\cI}$.  

Let us argue that $\sigma_{A'(\cI \hat{X}_{\cI})}$ has the properties required 
by Definition~\ref{def:wse}. First, observe that 
\begin{align*}
\sigma_{A'\hat{X}^n\hat{\Theta}^n\tilde{\Theta}^n}&=\sigma_{A'\hat{X}^n}\otimes\tau_{\sbin^n}\otimes\tau_{\sbin^n}\ ,
\end{align*}
since both $\hat{\Theta}^n$ and $\tilde{\Theta}^n$ are chosen uniformly and independently at random by the simulator and Bob, respectively. 
Since the set~$\cI$ consists of those indices where $\hat{\Theta}^n$ and $\tilde{\Theta}^n$ agree, we conclude that $\cI$ is uniform on the set of 
subsets of $[n]$, and independent of $A'\hat{\Theta}^n$. That is, the previous identity implies
\begin{align}
\sigma_{A'\hat{X}^n\cI}&=\sigma_{A'\hat{X}^n}\otimes\tau_{2^{[n]}}\ ,\label{eq:nec}
\end{align} 
as desired.  

It remains to prove that the state created during the real protocol equals this ideal state, that is, 
\begin{align}
\rho_{A'B}&=\sigma_{(\Theta^n T)(\cI \hat{X}_\cI)}\ .\label{eq:secprop}
\end{align}
\noindent To produce the state $\sigma_{(\Theta^n T)(\cI \hat{X}_\cI)}$, honest Bob (interacting with the simulator) measures all qubits in the bases~$\tilde \Theta^n$. Since we are only interested in $\hat{X}_\cI$, we could instead apply the first measurement and re-encoding (by the simulator) and the second measurement (by Bob) only on the qubits in $\cI$ without affecting the output. But since for all $i \in \cI$, we have $\hat{\Theta}_{i} = \tilde{\Theta}_{i}$, the re-encoding and the second measurement are always in the same basis, and can therefore be removed. 
Therefore, the state $\sigma_{\Theta^n T\cI \hat{X}_\cI}$ can also be produced in the following way: Let Alice output registers $(Q_A, \Theta^n, T)$. We first choose $\cI \subset [n]$ uniformly at random. Then, we measure all qubits in $\cI$ in bases~$\Theta_{\cI}$ to get~$\hat X_{\cI}$, and output registers $(\Theta^n, T, \cI, \hat X_{\cI})$. Since all qubits in the complement $\cI^c$ are discarded anyway, we can measure them in $\tilde \Theta_{\cI^c}$ without affecting the reduced state~$\sigma_{\Theta^nT\cI\hat{X}_{\cI}}$. But this exactly what happens in the real protocol producing the state~$\rho_{A'B}$, which implies Eq.~\eqref{eq:secprop}.

\end{proof}

\subsection{Application to concrete tensor product channels}

We examine the security parameters we can obtain for several well-known  channels. A simple example is the $d$-dimensional depolarizing channel defined in~\eqref{eq:depolarizingchannel},
which replaces the input state $\rho$ with the completely mixed state with probability $1-r$. Another simple example is the
one-qubit two-Pauli channel~\cite{king:unital} 
$$
\cN_{\rm Pauli}(\rho) \assign r \rho + \frac{1-r}{2} X \rho X + \frac{1-r}{2} Z \rho Z\ .
$$
Both these channels obey the \scp~\eqref{eq:strongconverseproperty} (see~\cite{rs:converse}), allowing us to obtain security of 
weak string erasure by Corollary~\ref{cor:WSEdishonestBob}.

For simplicity, we first consider the case 
where the storage rate is $\nu = 1$, that is, Bob's storage system is~$(\Complex^d)^{\otimes n}$, i.e., $n$ copies of a $d$-dimensional system, and his noise channel is $\cF=\cN^{\otimes n}$.
We first determine the values of $r$ that allow for a secure implementation of weak string erasure.
By Corollary~\ref{cor:WSEdishonestBob}, the capacity of the channel $\cN$ must be bounded by $C_\cN<\frac{1}{2}$.
The table given in Figure~\ref{fig:threshold} summarizes the relevant parameters.

\begin{figure*}[t!]
\begin{center}
\begin{tabular}{|c|c|c|c|}
\hline
Channel & Capacity  $C_\cN$ & Reference & Threshold\\
\hline
\hline
Qubit depolarizing& $1 + \frac{1+r}{2} \log \frac{1+r}{2} +  \frac{1-r}{2} \log \frac{1-r}{2}$ &\cite{king:depol} & $r \leq 0.77$\\[2mm]
\hline
Qutrit depolarizing& $\log 3 + \left(r + \frac{1-r}{3}\right) \log \left(r + \frac{1-r}{3}\right)
+ 2 \frac{1-r}{3} \log \frac{1-r}{3}$ &\cite{king:depol} & $r \leq 0.61$\\[2mm]
\hline
Two-Pauli& $1 - h\left(\frac{1 + \max(r,2r-1)}{2}\right)$ &\cite{king:unital} & $r \leq 0.77$\\[2mm]
\hline
\end{tabular}
\end{center}
\caption{A sufficient condition for achieving security (for storage rate $\nu=1$) is that the noise parameter $r$ lies below the threshold given above. This is equivalent to $C_\cN<\frac{1}{2}$.}
\label{fig:threshold}
\end{figure*}

\begin{figure}[t!]
\begin{center}
\hspace{-10ex}\scalebox{0.85}{
\begin{pspicture}(-1.0,0)(9.0,6.0)
\psset{unit=.7cm}
\psset{linewidth=.8pt}
\psset{labelsep=2.5pt}
\put(0,0){\epsfig{file=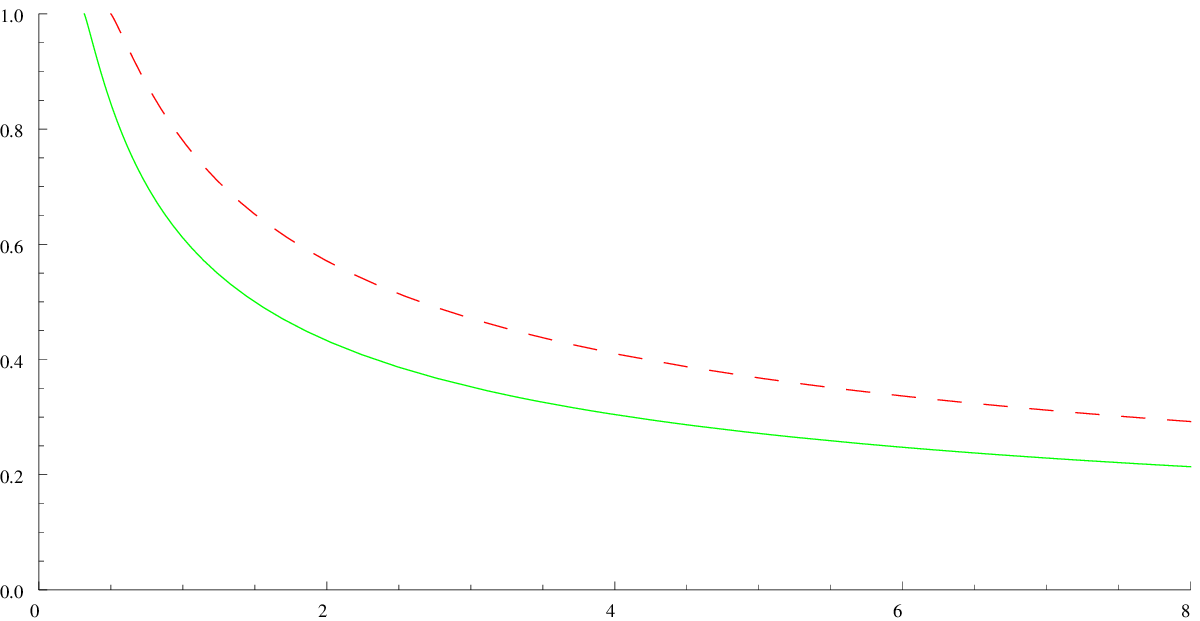,width=10cm}}
\put(1.0,2){secure realization}
\put(6,5){?}
\put(14.3,0.5){$\nu$}
\put(0.3,7.5){$r$}
\end{pspicture}
}
\caption{Tradeoff between $\nu$ and $r$: security can be obtained for the qutrit depolarizing channel below solid blue line and
the two-Pauli channel below the dashed red line. Note, however, that for the same storage rate the dimension of the storage system
is larger for the qutrit than for the qubit channel.}\label{fig:qutritDepol}
\end{center}
\end{figure}
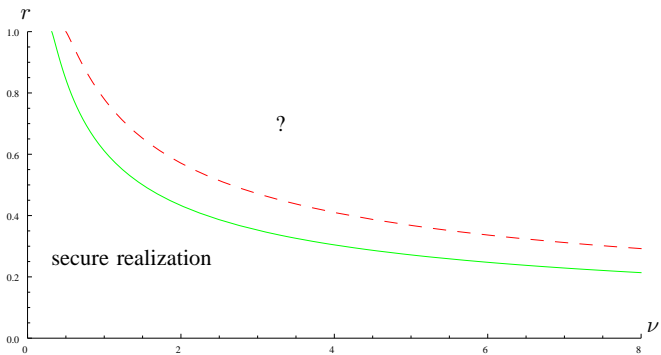

When allowing storage rates other than~$\nu=1$, we may again consider the regime where our proof provides security. Figure~\ref{fig:qutritDepol} 
examines this setting for the qutrit depolarizing channel and the two-Pauli channel, respectively.

\begin{figure}[t!]
\begin{center}
\hspace{-10ex}\scalebox{0.85}{
\mbox{\begin{pspicture}(-1.0,0)(9.0,5.8)
\psset{unit=.7cm}
\psset{linewidth=.8pt}
\psset{labelsep=2.5pt}
\put(0,0){
\epsfig{file=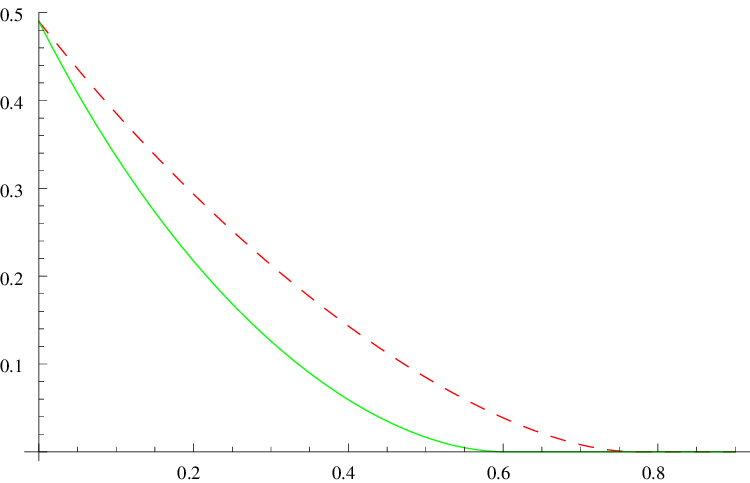,width=10cm}}
\put(14.6,0.5){$r$}
\put(0.6,9.2){$\lambda$}
\end{pspicture}}
}
\end{center}
\caption{The value of the min-entropy rate $\lambda$ for the qubit depolarizing channel (dashed red line) and the qutrit depolarizing channel (solid blue line) as a function of the noise parameter~$r$, for $\nu = 1$ and~$\delta = 0.01$.
Using qutrits means that the dimension of the overall storage system is higher, and we expect the resulting higher capacity to lead to a smaller
min-entropy rate~$\lambda$. Our analysis confirms this intuition.}
\label{fig:depolValue}
\end{figure}

To determine the exact security of the protocol, we need to compute the min-entropy rate 
$$
\lambda(\delta) = \nu \cdot \gamma^{\cN}\left(\frac{1/2 - \delta}{\nu}\right)\ ,
$$
as stated in Corollary~\ref{cor:tensorChannel}. For the class of channels~$\cN: \bop(\Complex^d) \rightarrow \bop(\Complex^d)$  considered in~\cite{rs:converse}, the 
strong converse property~\eqref{eq:strongconverseproperty} was shown to be satisifed with the function~$\gamma^{\cN}$ given by
\begin{align*}
\gamma^{\cN}(R) &= 
\max_{\alpha \geq 1} \frac{\alpha - 1}{\alpha}\left(R - \log d + S^{\rm min}_{\alpha}(\cN)\right)\ ,
\end{align*}
where $S^{\rm min}_{\alpha}(\cN)$ is the minimum output $\alpha$-R\'enyi-entropy of the channel. 
For the $d$-dimensional depolarizing channel (see~\eqref{eq:depolarizingchannel}) we may rewrite this expression~\cite{king:depol} as
$$
\gamma^{\cN}(R) = \max_{\alpha \geq 1} 
\left\{\begin{split}
\frac{\alpha-1}{\alpha}(R - \log d)\hspace{20ex} & \\
 - \frac{1}{\alpha}\log \left(
\begin{split}
&\left(r + \frac{1-r}{d}\right)^\alpha \\
&+ (d-1)\left(\frac{1-r}{d}\right)^\alpha
\end{split} \right)
\end{split} \right\}\ .
$$
Figure~\ref{fig:depolValue} shows how the min-entropy rate~$\lambda(\delta)$ relates to the noise
parameter $r$ for the qubit and qutrit depolarizing channels for a storage rate of $\nu = 1$ and error $\delta = 0.01$. 
The figure shows that the min-entropy rate we can achieve in our protocol is directly
related to the amount of noise in the storage.

\section{Bit commitment from weak string erasure\label{sec:bc} }

\subsection{Definition}
Informally, a standard \emph{commitment scheme} consists of a Commit
and an Open primitive between two parties Alice and Bob. First,
Alice and Bob execute the Commit primitive, where Alice
has input $Y^\ell \in \01^\ell$, and Bob has no input.
As output, Bob receives a notification that Alice has chosen an input $Y^\ell$.
Afterwards, they may execute the Open protocol, during which Bob either
accepts or rejects.
If both parties are honest, Bob always accepts and receives the value $Y^\ell$.
If Alice is dishonest, however, we still demand that Bob either 
outputs the correct
value of~$Y^\ell$ or rejects (binding).
If Bob is dishonest, he should not be able to gain any information about $Y^\ell$
before the Open protocol is executed (hiding).

\begin{figure*}[t!]
\begin{center}
\scalebox{1.0}{
\fbox{\begin{pspicture}(-5,0)(7,5)
\psset{unit=.7cm}
\psset{linewidth=.8pt}
\psset{labelsep=2.5pt}
\put(0,4){\psframe[linewidth=2pt,framearc=.3,fillstyle=solid,
fillcolor=lightgray](4,2)}
\put(1.1,4.8){Commit}
\put(0,5){\psline[linewidth=2pt]{->}(-2,0)}
\put(4,5){\psline[linewidth=2pt]{->}(2,0)}
\put(-3.5,6.3){Alice}
\put(6.5,6.3){Bob}
\put(-4.2,4.9){$C^\ell$}
\put(6.1,4.9){Committed}
\put(1.5,3){$Y^\ell \oplus C^\ell$}
\put(-3,2.7){\psline[linewidth=1pt,linestyle=dashed]{->}(9,0)}
\put(0,0){\psframe[linewidth=2pt,framearc=.3,fillstyle=solid,
fillcolor=lightgray](4,2)}
\put(1.3,0.8){Open}
\put(4,1){\psline[linewidth=2pt]{->}(2,0)}
\put(6.1,0.8){$\tilde{C}^\ell, F$}
\end{pspicture}}
}
\end{center}
\caption{Randomized string commitment:
Alice receives a random $C^\ell \in_R \01^\ell$ from Commit. During
the Open phase,
Bob outputs $\tilde{C}^\ell$ and~$F$. If both parties are honest, then $\tilde{C}^\ell=C^\ell$ and~$F=\textmath{accept}$. If Alice is dishonest, Bob 
outputs $F\in\{\textmath{accept},\textmath{reject}\}$, but
$\tilde{C}^\ell = C^\ell$ if  $F = \textmath{accept}$. To obtain a standard commitment,
Alice can send the extra message indicated by the
dashed line.}\label{fig:bc}
\end{figure*}
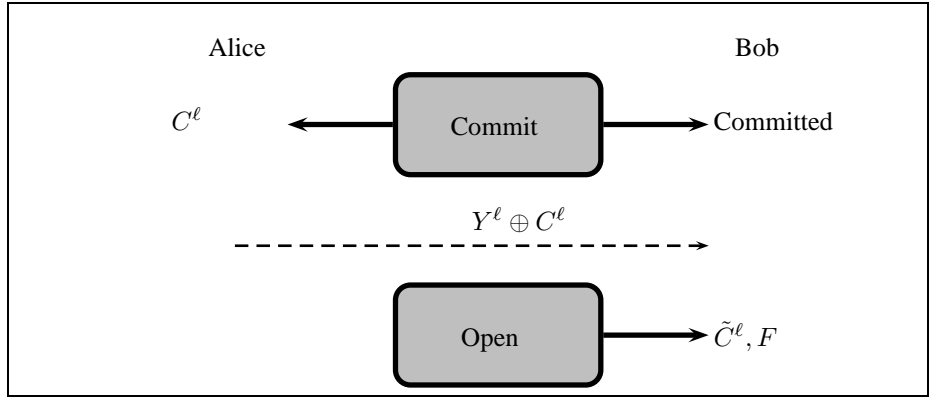

Here, we make use of a randomized version of a commitment as depicted
in Figure~\ref{fig:bc}. This simplifies both our definition, as well as the protocol.
Instead of inputting her own string $Y^\ell$, Alice now
\emph{receives} a random string $C^\ell$ from the Commit protocol.
Note that if Alice wants to commit to a value $Y^\ell$ of her choice,
she may simply
send the xor of her value with the random commitment $Y^\ell \oplus
C^\ell$ to Bob at the
end of the Commit protocol.

To give a more formal definition, note that we may write the Commit and
the Open protocol as CPTPMs $\cC_{AB}$ and $\cO_{AB}$ respectively,
consisting of the local actions
of honest Alice and Bob, together with any operations they may perform
on messages that are exchanged.
When both parties are honest, the output of the Commit protocol will
be a state $\cC_{AB}(\rho_{\rm in}) = \rho_{C^\ell UV}$
for some fixed input state $\rho_{\rm in}$, where
$C^\ell \in \01^\ell$ is the classical output of Alice, and $U$ and
$V$ are the internal states of
Alice and Bob respectively.
Clearly, if Alice is dishonest, she may not follow the protocol, and
we use $\cC_{A'B}$ to denote
the resulting map.
Note that $\cC_{A'B}$ may not have output~$C^\ell$, and we hence simply write
$\rho_{A'V}$ for the resulting output state, where $A'$ denotes the
register of a dishonest Alice.
Similarly, we use $\cC_{AB'}$ to denote the CPTPM corresponding to the
case where Bob is dishonest,
and write $\rho_{C^\ell UB'}$ for the resulting output state, where
$B'$ denotes the register of a dishonest Bob.

The Open protocol can be described similarly. If both parties are honest, the
map $\cO_{AB}: \cB(\hil_{UV}) \rightarrow
\cB(\hil_{\tilde{C}^\ell F})$ creates the state
$\eta_{C^\ell\tilde{C}^\ell F} := (\id_{C^{\ell}} \otimes \cO_{AB})
(\rho_{C^\ell UV})$, where
$\tilde{C}^\ell \in \{0,1\}^\ell$ and $F \in
\{\mbox{accept},\mbox{reject}\}$ is the
classical output of Bob.
Again, if Alice is dishonest, we write $\cO_{A'B}$ to denote the
resulting CPTPM with output~$\eta_{A''\tilde{C}^\ell F}$, and if Bob is dishonest, we write
$\cO_{AB'}$ for the resulting CPTPM with output
$\eta_{C^\ell B''}$. The following definition is similar to the one
given in~\cite{serge:new}, but slightly more general.

\begin{definition}
An {\em $(\ell,\eps)$-randomized string commitment scheme} is a protocol between
Alice and Bob satisfying the following properties:
\begin{enumerate}[]
\item{\em Correctness:}  If both parties are honest, then the ideal state
$\sigma_{C^\ell C^\ell F}$ is defined such that
\begin{enumerate}
\item The distribution of $C^\ell$ is uniform, and Bob accepts the
commitment:
$$
\sigma_{C^\ell F} =\tau_{\sbin^\ell} \otimes \ketbra{accept}{accept}\ .
$$
\item The joint state $\eta_{C^\ell \tilde{C}^\ell F}$ created by the
real protocol is $\eps$-close to the ideal state:
$$
\eta_{C^\ell \tilde{C}^\ell F} \approx_{\eps} \sigma_{C^\ell C^\ell F}\ ,
$$
where we identify $(A,B)$ with $(C^\ell,\tilde{C}^\ell F)$. 

\end{enumerate}
\item{\em Security for Alice ($\eps$-hiding):}
If Alice is honest, then for any joint state $\rho_{C^\ell B'}$
created by the Commit protocol, Bob does not learn $C^\ell$:
$$
\rho_{C^\ell B'}\approx_\eps \tau_{\{0,1\}^\ell} \otimes \rho_{B'}\ .
$$
\item{\em Security for Bob ($\eps$-binding):} If Bob is honest, then there
exists an ideal
cqq-state $\sigma_{C^\ell A'V}$ such that for all $\cO_{A'B}$:
\begin{enumerate}
\item Bob almost never accepts $\tilde{C}^\ell \neq C^\ell$:
\begin{align*}
\mbox{For } \psi_{C^\ell A''\tilde{C}^\ell F} = (\id_{C^\ell} \otimes
\cO_{A'B}) (\sigma_{C^\ell A'V})\ , \mbox{ we have }\\
 \Pr[C^\ell \neq \tilde{C}^\ell \mbox{ and } F =
\textmath{accept}] \leq \eps\ .\qquad\qquad 
\end{align*}
\item The joint state $\rho_{A'V}$ created by the real protocol is
$\eps$-close to the ideal state:
$$
\rho_{A'V}
\approx_\eps
\sigma_{A'V}\ .
$$
\end{enumerate}
\end{enumerate}
\end{definition}

\subsection{Protocol}

Let $\eps' >0$. To construct our protocol based on weak string erasure, we will need a binary
$(n,k,d)$-linear code $\cC \subseteq \{0,1\}^n$, i.e., a linear code
with $2^{k}$ elements
and minimal distance $d:=2\log1/\eps'$.
Let $\Syn : \{0,1\}^n \rightarrow \{0,1\}^{n-k}$ be a function that
outputs a parity-check syndrome for the code $\cC$.
Let $\Ext: \01^n \times \cR \rightarrow \01^\ell$
be a $2$-universal hash function as defined in
Section~\ref{sec:privacyamplification}

\begin{protocol}{2a}{Commit}{Inputs: none. Outputs: $c^\ell \in
\{0,1\}^\ell$ to Alice.}
\item {\bf Alice and Bob:} Execute $(n,\lambda,\eps)$-WSE. Alice gets
$x^n \in \{0,1\}^n$, and Bob gets $\cI \subset [n]$ and $s =x_\cI$.
\item {\bf Alice:} Chooses $r \in_R \cR$ and sends
$r$ and $w := \Syn(x^n)$ to Bob.
\item {\bf Alice:} Outputs $c^\ell := \Ext(x^n,r)$ and stores $x^n$. Bob stores
$(r,w,\cI,s)$.
\end{protocol}

\begin{protocol}{2b}{Open}{Inputs: none. Outputs: $\tilde{c}^\ell \in \{0,1\}^\ell$ and 
$f \in  \{\textmath{accept},\textmath{reject}\}$ to Bob.}
\item {\bf Alice:} Sends $x^n$ to Bob.
\item {\bf Bob:} If $s \neq x_\cI$ or $w \neq \Syn(x^n)$, then he outputs
$\tilde{c}^\ell := 0^\ell$ and $f := \textmath{reject}$. Otherwise, he
outputs $\tilde{c}^\ell := \Ext(x^n,r)$ and $f := \textmath{accept}$.
\end{protocol}
 
Our main claim of this section is the following.
\begin{theorem}[String commitment] \label{thm:WECtoBC}
The pair (2a, 2b) of protocols (Commit,Open) is an $(\lambda n - (n-k) - 2
\log 1/\eps', 2 \eps + \eps')$-randomized
string commitment scheme based on one instance of $(n,\lambda,\eps)$-WSE.
\end{theorem}

The length $\ell := \lambda n - (n-k) - 2 \log 1/\eps'$ of the commitment depends on our choice
of code $\cC$. 
Since we require that $\ell > 0$, we need $n-k$ to be small compared
to $n$, which means that we need codes for which $k / n  \rightarrow 1$
for $n \rightarrow \infty$.
A simple construction of codes that satisfy this can be based on
\emph{Reed-Solomon codes} \cite{ReedSolomon:1960} over the field
$GF(2^m)$, which are
$(2^m - 1,2^m - d, d)$-linear codes. We can convert these codes into
binary $( (2^m - 1)m , (2^m - d)m , d)$-linear codes by simply mapping
each field element to $m$ bits. For $n := (2^m - 1)m$, we have $n-k =
(d-1) m \leq d (\log n-1)$, since $n \geq 2 \cdot 2^m $ whenever $m \geq
3$. Therefore, with these codes we can achieve
$\ell \geq \lambda n - 2 \log n \log 1/\eps$, i.e., our commitment rate is
roughly~$\lambda$.

\subsection{Security proof}

We again show security for Alice and Bob individually. Recall that
if Bob is dishonest, our goal is to show
that his information about $C^\ell$ is negligible.
The intuition behind this proof is that weak string erasure ensures that Bob's information about the string $X^n$ is limited.
Via privacy amplification we then obtain that his information about $C^\ell$, which is the output of a $2$-universal
hash function applied to $X^n$, is neglible.

\begin{lemma}[Security for Alice] 
\label{lem:bc-hiding}
The pair of protocols (Commit,Open) is $(2 \eps + \eps')$-hiding.
\end{lemma}

\begin{proof}
Let $\rho_{X^n B'}$ the cq-state created by the execution of WSE.
From the properties of WSE it follows that there exists a state
$\sigma_{X^n B'}$ such that $\hmin(X^n|B')_\sigma \geq \lambda n$
and $\rho_{X^n B'} \approx_\eps \sigma_{X^nB'}$. This implies that
\[\hmin^{\eps}(X^n|B')_\rho \geq \lambda n\;.\]
By the chain rule (see~\eqref{eq:chainrule}), we get
\[\hmin^{\eps}(X^n|B'\Syn(X^n))_\rho \geq \lambda n - (n-k) = \ell + 2
\log 1/\eps'\;.\]
Using privacy amplification (Theorem \ref{thm:PA}), we then get that
$$
\frac{1}{2} \| \rho_{C^\ell B'} - \tau_{\01^{\ell}} \otimes
\rho_{B'}\|_1 \leq 2\eps + 2^{- \frac12 \cdot 2 \log 1/\eps' - 1} \leq
2 \eps + \eps'\;,
$$
as promised.
\end{proof}

To show security for honest Bob, we need the following property of linear codes.
Note that the function $\Syn$ is linear, i.e., for all codewords $x^n$ and $\bar{x}^n$, we
have $\Syn(x^n \oplus \bar{x}^n) =  \Syn(x^n) \oplus \Syn(\bar{x}^n)$.
Therefore, for any $x^n$ and $\bar{x}^n$ with $x^n \neq \bar{x}^n$ and $\Syn(x^n) =
\Syn(\bar{x}^n)$, we have that the string $\Syn(x^n \oplus \bar{x}^n) \in \{0,1\}^{n-k}$ is the all zero string~$0^{n-k}$. From this it follows that $x^n\oplus \bar{x}^n$ is a codeword different from $0^{n}$. Since all codewords except $0^{n}$
have weight at least $d$, 
it follows that~$x^n$ and~$\bar{x}^n$ have distance at least~$d$.

The intuition behind the following proof is the observation that weak string erasures ensures that Bob knows the substring~$\hat{X}_{\cI}$ of a string~$\hat{X}$. The properties of the error-correcting code limit the set of strings~$X^n$ consistent with this substring and the given syndrome~$W$; this implies that Alice  will be detected with 
high probability if she attempts to cheat.

\begin{lemma}[Security for Bob]\label{lem:bc-binding}
The pair of protocols (Commit,Open) is $\eps$-binding.
\end{lemma}

\begin{proof}
Let $\rho_{A' B}$ be the state shared by Alice and Bob after the
execution of WSE.
From the properties of WSE it follows that there exists a state
$\sigma_{A' \hat{X}^n \cI} = \sigma_{A' \hat{X}^n} \otimes
\tau_{2^{[n]}}$ such that
$\rho_{A' B} = \sigma_{A' (\cI \hat{X}_\cI)}$, where $B = (\cI \hat{X}_\cI)$.
Let $\bar{X}^n$ be the
closest string to $\hat{X}^n$ that satisfies $\Syn(\bar{X}^n) =
W$, and let $C^\ell := \Ext(\bar{X}^n,R)$. We will now show that the
state $\sigma_{C^\ell A' (R W \cI S)}$ created during the Commit protocol
satisfies the binding
condition.

First of all, note that if Alice sends $X^n = \bar{X}^n$, then Bob
outputs $\tilde{C}^\ell = C^\ell$.
It thus remains to analyze the case of $X^n \neq \bar{X}^n$. Note that
we may write
\begin{align*}
&\Pr [C^\ell \neq \tilde{C}^\ell \mbox{ and } F = \textmath{accept}]
\\[2mm]
&=\hspace{-3ex}\Pr_{\substack{R,X^n,\bar{X}^n\\ \Syn(X^n) \neq
\Syn(\bar{X}^n)}}
[\Ext(\bar{X}^n,R) \neq \Ext(X^n,R) \mbox{ and } F = \textmath{accept}]\\
&\ \  +\hspace{-3ex}
\Pr_{\substack{R,X^n,\bar{X}^n\\ \Syn(X^n) = \Syn(\bar{X}^n)}}
[\Ext(\bar{X}^n,R) \neq \Ext(X^n,R) \mbox{ and } F = \textmath{accept}]\\[2mm]
&=\hspace{-3ex}
\Pr_{\substack{R,X^n,\bar{X}^n\\ \Syn(X^n) = \Syn(\bar{X}^n)}}
[\Ext(\bar{X}^n,R) \neq \Ext(X^n,R) \mbox{ and } F = \textmath{accept}]
\end{align*}
where the last equality follows from the fact that Bob always rejects
if $\Syn(X^n) \neq \Syn(\bar{X}^n)$.

We now show that the remaining term is small. Note that if $\Syn(X^n)
= \Syn(\bar{X}^n)$,
and $X^n \neq \bar{X^n}$, the distance between $X^n$ and $\bar{X}^n$
is at least $d$.
We also know that for our choice of~$\bar{X}^n$, the distance
between $\bar{X}^n$ and
$\hat{X}^n$ is at most $d/2$. Hence, $X^n$ has distance at least $d/2$
to~$\hat{X}^n$.
Since Alice does not know
$\cI$ and every $i \in [n]$ is in $\cI$ with probability $\frac12$,
Bob accepts with probability at most $\eps = 2^{-d/2}$. Hence, we obtain
$$
\Pr[C^\ell \neq \tilde{C}^\ell \mbox{ and } F = \textmath{accept}] \leq \eps'\ ,
$$
as promised.
\end{proof}

It remains to show that the protocol is correct. This follows
essentially from the properties of weak string erasure. However, we
still need to demonstrate that the state we obtain from weak string
erasure has~$C^\ell$ close to uniform.

\begin{lemma}[Correctness] 
\label{lem:bc-corr}
The pair of protocols (Commit,Open) satisfies correctness with an
error of at most $2 \eps + \eps'$.
\end{lemma}

\begin{proof}
Let $\eta_{C^\ell \tilde{C}^\ell}$ be the state at the end of the
protocol. It follows
directly from the properties of WSE that $\eta_{C^\ell \tilde{C}^\ell}
= \eta_{C^\ell C^\ell}$.
It remains to show that this state is close to the ideal state
$\sigma_{C^\ell C^\ell}$.
By the same arguments as in Lemma \ref{lem:bc-hiding} it follows that
$\frac{1}{2}\|\eta_{C^\ell} - \sigma_{C^\ell}\|_1 \leq 2 \eps + \eps'$.
Hence, we also have $\frac{1}{2}\|\eta_{C^\ell C^\ell} -
\sigma_{C^\ell C^\ell} \|_1 \leq 2 \eps + \eps'$.
\end{proof}

\section{1-2 oblivious transfer from weak string erasure\label{sec:ot}}

\subsection{Definition}\label{sec:otdef}
We now show how to obtain 1-2 oblivious transfer given access to weak
string erasure. Usually, one considers a non-randomized version of 1-2
oblivious transfer, in which Alice has two inputs
$Y_0^\ell, Y_1^\ell \in \01^\ell$,
and Bob has as input a choice bit $D \in \01$. At the end of the
protocol Bob receives $Y_D^\ell$,
and Alice receives no output. The protocol is considered
secure if the parties do not gain any information beyond this
specification, that is, Alice does not learn $D$ and there exists some
input $Y_{1-D}^\ell$
about which Bob remains ignorant.

Here, we again make use of fully randomized
oblivious transfer. Fully randomized oblivious transfer takes no inputs,
and outputs two strings $S_0^\ell,S_1^\ell \in \01^\ell$
to Alice, and a choice bit $C \in \01$ and $S_C^\ell$ to Bob. Security
means that if Alice is dishonest,
she should not learn anything about $C$. Similar to weak string
erasure, we also demand
that two strings $S_0^\ell$ and $S_1^\ell$ are created by the protocol.
Intuitively, this ensures that just like in a classical protocol, we can again
think of the protocol as being
completed once Alice and Bob have exchanged their final message.
If Bob is dishonest, we demand that there exists
some random variable $C$
such that Bob is entirely ignorant about $S_{1-C}^\ell$. That is, he may
learn at most one of the two
strings which are generated.

\begin{figure*}[t!]
\begin{center}
\scalebox{1.0}{
\fbox{\begin{pspicture}(-5,0)(7,4)
\psset{unit=.7cm}
\psset{linewidth=.8pt}
\psset{labelsep=2.5pt}
\put(0,3){\psframe[linewidth=2pt,framearc=.3,fillstyle=solid,
fillcolor=lightgray](4,2)}
\put(1.3,3.8){FROT}
\put(0,4){\psline[linewidth=2pt]{->}(-2,0)}
\put(4,4){\psline[linewidth=2pt]{->}(2,0)}
\put(-3.5,5){Alice}
\put(6.5,5){Bob}
\put(-5.5,3.9){$S_0^\ell,S_1^\ell \in \01^\ell$}
\put(6.3,3.9){$C \in \01,S_C^\ell$}
\put(0.5,2){$M = C \oplus D$}
\put(-3,1.7){\psline[linewidth=1pt,linestyle=dashed]{<-}(10,0)}
\put(0,0.5){$S_0^\ell \oplus Y_{M}^\ell, S_1^\ell \oplus Y_{1-M}^\ell$}
\put(-3,0.2){\psline[linewidth=1pt,linestyle=dashed]{->}(10,0)}
\end{pspicture}}
}
\caption{Fully randomized 1-2-oblivious transfer when Alice and Bob
are honest. Intuitively, if one of the parties is dishonest, he/she
should not be able to obtain more information from the primitive as
depicted above. The dashed messages are exchanged to obtain
non-randomized oblivious transfer from FROT.}\label{fig:frot}
\end{center}
\end{figure*}
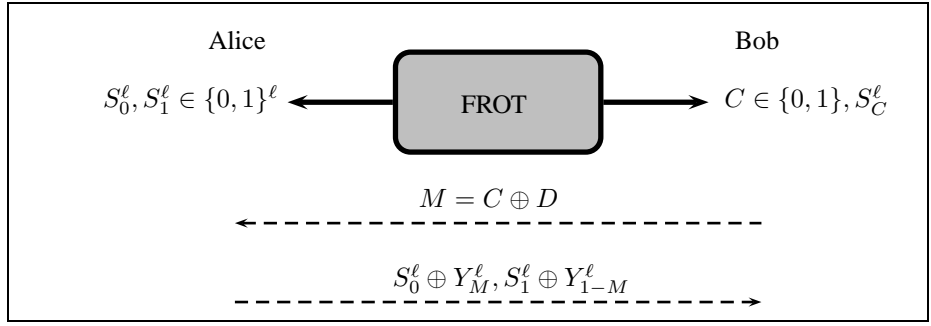

Fully randomized oblivious transfer can easily be converted into ``standard'' oblivious
transfer as depicted in Figure~\ref{fig:frot}
using the protocol presented in \cite{BBCS92} (see also
\cite{Beaver95}). To obtain non-randomized 1-2 oblivious transfer,
Bob sends Alice a message indicating whether $C = D$.
Note that since Alice does not know~$C$, she also does not know anything about~$D$. If $C=D$, Alice sends Bob $Y_0^\ell \oplus S_0^\ell$, and
$Y_1^\ell \oplus S_1^\ell$, otherwise she sends $Y_0^\ell \oplus
S_1^\ell$ and $Y_1^\ell \oplus S_0^\ell$.
Clearly, if Bob does not learn anything about~$S_{1-C}^\ell$,
he can learn at most one of $Y_0^\ell$ and $Y_1^\ell$~\cite{BBCS92,Beaver95}.

We now provide a more formal definition, which is very similar to the
definitions in \cite{serge:new,fs:compose}.

\begin{definition}
An {\em $(\ell,\eps)$-fully randomized oblivious transfer (FROT) scheme} is a
protocol between Alice and Bob satisfying the following:
\begin{enumerate}
\item{\em Correctness:} If both parties are honest, then the ideal state
$\sigma_{S_0^\ell S_1^\ell C S_C^\ell}$ is defined such that
\begin{enumerate}
\item The distribution over $S_0^\ell$, $S_1^\ell$ and $C$ is uniform:
$$
\sigma_{S_0^\ell S_1^\ell C}=\tau_{\sbin^\ell} \otimes
\tau_{\sbin^\ell} \otimes \tau_{\sbin}\ .
$$
\item The real state $\rho_{S_0^\ell S_1^\ell C Y^\ell}$ created
during the protocol is $\eps$-close to the ideal state:
\begin{align}
\rho_{S_0^\ell S_1^\ell C Y^\ell} \approx_\eps \sigma_{S_0^\ell
S_1^\ell C S_C^\ell}\ ,
\end{align}
where we identify $A=(S_0^\ell,S_1^\ell)$ and $B=(C,Y^\ell)$.
\end{enumerate}
\item{\em Security for Alice:} If Alice is honest, then there exists an
ideal state $\sigma_{S_0^\ell S_1^\ell B'C}$,
where $C$ is a random variable on $\sbin$, such that
\begin{enumerate}
\item Bob is ignorant about $S_{1-C}^\ell$:
$$
\sigma_{S_{1-C}^\ell S_C^\ell B'C} \approx_\eps
\tau_{\sbin^\ell}\otimes\sigma_{S_C^\ell B'C}\ .
$$
\item The real state $\rho_{S_0^\ell S_1^\ell B'}$ created during the
protocol is $\eps$-close to the ideal state:
$$
\rho_{S_0^\ell S_1^\ell B'} \approx_\eps \sigma_{S_0^\ell S_1^\ell B'}\ .
$$
\end{enumerate}

\item{\em Security for Bob:} If Bob is honest, then there exists an ideal state
$\sigma_{A'S_0^\ell S_1^\ell C}$ such that
\begin{enumerate}
\item Alice is ignorant about $C$:
$$
\sigma_{A'S_0^\ell S_1^\ell C} = \sigma_{A'S_0^\ell
S_1^\ell}\otimes\tau_{\sbin}\ .
$$
\item The real state $\rho_{A'C Y^\ell}$ created during the protocol
is $\eps$-close to the ideal state:
$$
\rho_{A'C Y^\ell} \approx_\eps \sigma_{A'CS_C^\ell}\ ,
$$
where we identify $B = (C,Y^\ell)$.
\end{enumerate}
\end{enumerate}
\end{definition}

Again, we allow the protocol implementing this primitive to abort, but demand that the security 
conditions are satisfied if the protocol does not abort.

\subsection{Protocol}
We now show how to obtain a fully randomized oblivious transfer given access to weak string erasure.

As in Section~\ref{sec:wse}, honest players never abort the protocol. If the dishonest player refuses to send correctly formed messages, the honest player chooses the messages himself. Note that we require the same also from the interactive hashing protocol: If one player aborts it, the other terminates the interaction and proceeds to simulate the remainder of the protocol himself. Indeed, this is needed to satisfy Lemma~\ref{lem:IH}, which does not deal with aborts. By inspection of the protocols, it is easy to see that the honest player can indeed simulate all the other player's messages in this way.

To obtain some intuition 
for the actual protocol, consider the following na\"ive protocol, which we only state informally. It makes use of a $2$-universal hash function $\Ext:\sbin^{n/4}\times\cR\rightarrow \sbin^\ell$.

\begin{protocol}{3'}{Na\"ive Protocol (informal)}{Outputs: $(s_0^\ell,s_1^\ell) \in
\{0,1\}^{\ell} \times \01^\ell$ to Alice, and $(c,y^\ell) \in \{0,1\} \times \{0,1\}^\ell$
to Bob}

\item {\bf Alice and Bob: }  Execute WSE. Alice gets a string $x^n \in \{0,1\}^n$, Bob a set $\cI \subset [n]$
and a string $s = x_\cI$.
 If $|\cI| < n/4$, Bob randomly adds elements to $\cI$ and pads the corresponding positions in~$s$ with $0$s.
Otherwise, he
randomly truncates $\cI$ to size~$n/4$, and deletes the
corresponding values in~$s$.

\item {\bf Alice and Bob:} Execute interactive hashing with Bob's input~$w$ equal to a description of~$\cI=\Enc(w)$. 
Interpret the outputs $w_0$ and $w_1$ as descriptions of subsets $\cI_0$ and $\cI_1$ of~$[n]$.
\item {\bf Alice:} Chooses $r_{0}, r_{1} \in_R \cR$ and sends them to Bob.
\item {\bf Alice:} Outputs $(s_0^\ell,s_1^\ell):=$\\ \mbox{\hspace{20ex} $(\Ext(x_{\cI_0},r_0),\Ext(x_{\cI_1},r_1))$}.
\item {\bf Bob: } Computes $c\in\sbin$ with $\cI=\cI_c$, and $x_{\cI}$ from $s$. He outputs $(c,y^\ell) \assign
(c,\Ext(s,r_c))$.
\end{protocol}

For now, let us neglect the fact that the outputs of interactive hashing are strings, and 
assume that the subset~$\cI_{1-c}$ generated by the interactive hashing protocol 
is uniformly distributed over subsets of size~$n/4$ not equal to~$\cI$. 
The string~$x_{\cI_{1-c}}$ is then obtained by sampling from the string~$x^n$, which by the definition of weak string
erasure has high min-entropy. We therefore expect the value $s_{1-c}^\ell$ to be uniform and independent of Bob's view. 
This should imply security for Alice, whereas security for Bob immediately follows from the properties of interactive hashing.

\begin{protocol}{3}{WSE-to-FROT}{
Parameters: Integers $n,\beta$ such that  $m:=n/\beta$ 
is a multiple of~$4$.
 Set~$t := \lfloor \log  \binom{n}{n/4} \rfloor$.
Outputs: $(s_0^\ell,s_1^\ell) \in
\{0,1\}^{\ell} \times \01^\ell$ to Alice, and $(c,y^\ell) \in \{0,1\} \times \{0,1\}^\ell$
to Bob }
\item {\bf Alice and Bob: }  Execute $(n,\lambda,\eps)$-WSE.
Alice gets a string $x^n \in \{0,1\}^n$, Bob a set $\cI \subset [n]$
and a string $s = x_\cI$. If $|\cI| < n/4$, Bob randomly adds elements to $\cI$ and pads the corresponding positions in~$s$ with~$0$s.
Otherwise, he
randomly truncates $\cI$ to the size $n/4$, and deletes the
corresponding values in $s$. 
\item {\bf Bob: }
\begin{enumerate}
\item Randomly chooses a string $w^t\in_R \{0,1\}^t$ corresponding to an encoding  of a subset $\Enc(w^t)$ of  $[n]$ with $n/4$ elements.
\item   He randomly chooses a permutation $\pi:[n] \rightarrow [n]$ of the entries of $x$, such that he knows $\pi(x)_{\Enc(w^t)}$ (that is, these bits are permutation of the bits of~$s$). Formally, $\pi$ is uniform over permutations satisfying the following condition: for all $j \in [n]$ and $j' := \pi (j)$, we have $j \in \cI  \Leftrightarrow j' \in \Enc(w^t)$. 
\item Bob sends $\pi$ to Alice.
\end{enumerate}
\item {\bf Alice and Bob:} Execute interactive hashing with Bob's input
equal to $w^t$. They obtain $w_0^t,w_1^t \in \{0,1\}^t$ with
$w^t\in\{w_0^t,w_1^t\}$. 
\item {\bf Alice: } Chooses $r_{0}, r_{1} \in_R \cR$ and sends them to Bob.
\item {\bf Alice: } Outputs $(s_0^\ell,s_1^\ell) := (\Ext(\pi(x)_{\Enc(w_0^t)},r_0),
\Ext(\pi(x)_{\Enc(w_1^t)},r_1))$.
\item {\bf Bob: } Computes $c$, where $w^t=w_c^t$, and  $\pi(x)_{\Enc(w^t)}$ from $s$. He outputs $(c,y^\ell) :=(c,\Ext(\pi(x)_{\Enc(w^t)},r_c))$. 
\end{protocol}

To use interactive hashing in conjunction with subsets, the actual protocol needs an encoding of  subsets~$\Enc:\sbin^t\rightarrow \cT$, where~$\cT$ is the set of all subsets of $[n]$ of size~$n/4$.  Here we choose~$t$ such that $2^t\leq \binom{n}{n/4}\leq 2\cdot 2^t$, and an injective encoding $\Enc:\sbin^t\rightarrow\cT$, i.e., no two strings are mapped to the same subset. Note that this means that not all possible subsets are encoded, but at least half of them. We refer to~\cite{ding2,savvides:diss} for details on how to obtain such an encoding.
 Note that since not every subset has an encoding, we cannot simply take $w^t := \Enc^{-1}(\cI)$. To solve this problem, we first choose  a~$w^t$ uniformly at random, and then choose a random permutation $\pi$ such that Bob knows the subset encoded by $w^t$ in $\pi(x)$.

\begin{theorem}[Oblivious transfer]
For any constant $\omega \geq 2$ and $\beta \geq \max\{67,256 \omega^2/\lambda^2\}$, the protocol WSE-to-FROT implements an $(\ell,43 \cdot 2^{-\frac{\lambda^2}{512 \omega^2 \beta} n} + 2 \eps)$-FROT from  one instance of of $(n, \lambda,
\eps)$-WSE, where
$\ell := \left \lfloor \left(\left(\frac{\omega-1}{\omega}\right)\frac{\lambda}{8}-\frac{\lambda^2}{512 \omega^2\beta}\right)n - \frac{1}{2} \right \rfloor$.
\end{theorem}

Since this work is a proof of principle, we may choose $\omega=2$. However, if we were to look
at a more practical setting, choosing other values of $\omega$ can be beneficial.

\subsection{Security proof}

We first show that the protocol is secure against a cheating Alice.
Intuitively, the properties of weak string erasure ensure that Alice does not know which bits~$x_{\cI}$ of $x^n$ are known to Bob, that is, she is ignorant about the index set~$\cI$.  This implies that essentially any partition of the bits is consistent with Alice's view. In particular, she does  not gain much information from the particular partition chosen by Bob. Finally, the properties of interactive hashing ensure that she  cannot gain much information 
about which  of the two final strings is known to Bob.

\begin{lemma}[Security for Bob] 
\label{lem:FROT-secBob}
Protocol WSE-to-FROT satisfies security for Bob.
\end{lemma}

\begin{proof}
Let $\tilde
\rho_{A'' C Y^\ell }$ denote the joint state at the end of the protocol, 
where $A''$ is the quantum output of a malicious
Alice and $(C,Y^\ell)$ is the classical output of an honest Bob. We
 construct an ideal state $\tilde \sigma_{A'' W_0^\ell W_1^\ell C}
= \tilde \sigma_{A'' W_0^\ell W_1^\ell} \otimes \tau_{\{0,1\}}$ that
satisfies
$\tilde \rho_{A'' C Y^\ell } = \tilde \sigma_{A'' C W_C^\ell }$.

First, we divide a malicious Alice into two parts. The first part
interacts with Bob in the WSE protocol, after which the state shared
by Alice and Bob is
$\rho_{A' X_{\cI} \cI}$. 
From the
properties of WSE it follows that there exists an ideal state
$\sigma_{A' \hat{X}^n \cI \hat{X}_{\cI}}$ such that the reduced state satisfies
$\rho_{A' X_{\cI} \cI} = \sigma_{A' \hat{X}_{\cI} \cI}$.

The second part of Alice takes $A'$ as input
and interacts with Bob in the rest of the protocol.
To analyze the resulting joint output state~$\tilde{\rho}_{A''CY^\ell}$, we can use the properties of weak string erasure, and 
 let the second part of Alice interact with honest Bob
starting from the state $\sigma_{A' \hat{X}^n \cI}$. 
The
protocol outputs a state
$\tilde \sigma_{A'' \hat{X}^n C Y^\ell M}$, where $M$ denotes all
classical communication during the protocol. Note that the values
$\Pi$, $W_0^t$, $W_1^t$, $R_0$ and $R_1$ can be computed from $M$. Let
  $S_i^\ell := \Ext(\Pi(\hat X^n)_{\Enc(W_i^t)},R_i)$ 
for $i \in \{0,1\}$. 
We obtain the state $\tilde \sigma_{A'' S_0^\ell S_1^\ell C Y^\ell }$ by taking the partial trace of
$\tilde \sigma_{A'' S_0^\ell S_1^\ell \hat{X}^n C Y^\ell  M}$.
From the construction of this state and the fact that $\rho_{A'
X_{\cI} \cI} = \sigma_{A' \hat{X}_{\cI} \cI}$ it follows directly that
$\tilde \rho_{A'' C Y^\ell } = \tilde \sigma_{A'' C Y^\ell }$
and
$\tilde \sigma_{A''S_0^\ell S_1^\ell C Y^\ell } = \tilde \sigma_{A''S_0^\ell
S_1^\ell C S_C^\ell }$. Hence 
\[
 \tilde \rho_{A'' C Y^\ell }  = \tilde \sigma_{A'' C S_C^\ell }\;.
\]

It remains to be shown that Alice does not learn anything about $C$, that is,
$\tilde \sigma_{A'' S_0^\ell S_1^\ell C} = \tilde \sigma_{A'' S_0^\ell
S_1^\ell} \otimes \tau_{\{0,1\}}$.
From the properties of WSE it follows that
$\sigma_{A'\hat{X}^n\cI}=\sigma_{A'\hat{X}^n}\otimes\tau_{2^{[n]}}$.
Since Bob randomly  truncates/extends~$\cI$ such that $| \cI | = n/4$, the resulting set~$\cI$ is also uniformly distributed over all
subsets of size~$n/4$ and independent of~$A'$.
Hence, conditioned on any fixed $W^t = w^t$,
the permutation $\Pi$ is uniform and independent of $A'$.
It follows that the string $W^t$ is also uniform and independent of $A'$
and $\Pi$. From the properties of interactive hashing we are guaranteed that
 $C$ is uniform and independent of Alice's view afterwards, and hence,
 \[\tilde \sigma_{A'' S_0^\ell S_1^\ell C} = \tilde \sigma_{A''
S_0^\ell S_1^\ell} \otimes \tau_{\{0,1\}}\;.\]
\end{proof}

Second, we show that the protocol is secure against a cheating Bob. We again first give an intuitive argument.
We have from weak string erasure that Bob gains only a limited amount of information about the string~$X^n$.
The properties of interactive hashing ensure that Bob has very little
control over one of the  subsets of bits chosen by the interactive hashing. Therefore, by the results on min-entropy sampling, Bob only has limited information about these bits of~$X^n$. 
Privacy amplification can then be used to turn 
this into almost complete ignorance.
\begin{lemma}[Security for Alice]\label{lem:FROT-secAlice}
Protocol WSE-to-FROT satisfies security for Alice with an error of
\[41 \cdot 2^{-\frac{\lambda^2}{512 \omega^2\beta} n} + 2\varepsilon\;.\]
\end{lemma}

\begin{proof}
Let $\rho_{X^n B'}$ be the cq-state created by the execution of WSE.
From the properties of WSE it follows that there exists a state
$\sigma_{X^n B'}$ such that $\hmin(X^n|B')_\sigma \geq \lambda n$
and $\rho_{X^n B} \approx_\eps \sigma_{X^n B'}$, which implies that 
\[ \hmin^{\eps}(X^n | B')_\rho \geq \lambda n\;.\]
Since the permutation $\Pi$ is chosen by Bob based on his quantum information~$B'$, it follows from (\ref{eq:minentropymonotonicity}) that
\[\hmin^{\eps}(\Pi(X^n) | \Pi B'')_\rho = \hmin^{\eps}(X^n | \Pi B'')_\rho \geq \hmin^{\eps}(X^n | B')_\rho\;,\]
where $B''$ is Bob's part of the
shared quantum state after he has sent $\Pi$ to Alice.

Recall that our goal is to show that Bob has high min-entropy about the string $X^n$ 
restricted to one of the subsets generated by the interactive hashing protocol.
Our first step is to count the subsets which are bad for Alice in the sense
that Bob has a lot of information about~$X^n$. We then show that
the probability that both sets chosen via the interactive hashing primitive
lie in the bad set is exponentially small in $n$.

With
Lemma~\ref{lem:samplingmodified}, we conclude that for the uniform\footnote{Note that, in the protocol, we do not
actually sample from the uniform distribution over subsets; the
bound~\eqref{eq:boundsampler} is merely used in a counting argument
here to establish that the number of ``bad'' subsets is limited,
cf.~\eqref{eq:badsubsetsex} below.} distribution over subsets
$\cS\subset [n]=[\beta m]$ of
size $\beta m/4=|\cS|$
\begin{align}
\Pr_{\cS} \left [ \hmin^{\varepsilon+4\delta}(\Pi(X^n)_\cS|\cS \Pi B'')<\left(\frac{\omega-1}{\omega}\right)\frac{\lambda n}{4}\right ] \leq {\delta}^2\
,\label{eq:boundsampler}
\end{align}
where $\delta=2^{-m \lambda^2/(512 \omega^2)}$.
Let $\bad$ be the set of all
subsets of size $\beta m/4$ that result in small min-entropy, i.e.,
\[\bad := \left \{ \cS\ \Big|\  
\begin{split}
\cS\subset [\beta m]\ , |\cS|=\frac{\beta m}{4}
\textrm{ and }\hspace{5ex} &\\
\hmin^{\varepsilon+4\delta}(\Pi(X^n)_\cS|\cS \Pi B'') < 
\left(\frac{\omega-1}{\omega}\right)\frac{\lambda n}{4}\end{split} \right \}\;. \]
Since we have considered the uniform distribution over all subsets of~$[\beta m]$ of
size $\beta m/4$, we conclude from~\eqref{eq:boundsampler} that
\begin{align}
\begin{matrix}
|\{w^t\in\sbin^t\ |\ \Enc(w^t)\in \bad\}|&\leq & |\bad|\\
&\leq &\binom{\beta m}{\beta m/4} \delta^2\\
&\leq &2\cdot 2^t \delta^2\ .
\end{matrix}
\label{eq:badsubsetsex}
\end{align}
In the first inequality, we have used the fact that $\Enc$ is injective, i.e.,
every element in the image has exactly one preimage. In the last
inequality, we used the fact that~$\binom{\beta m}{\beta m/4}\leq 2\cdot
2^t$. By the third property of the interactive hashing, we conclude
that
\begin{align}
\begin{matrix}
\Pr\left[ \Enc(W_0^t) \in \bad\textrm{ and } \Enc(W_1^t) \in \bad\right]&\leq&
16\frac{2\cdot 2^t \delta^2}{ 2^t}\\
& \leq & 32\delta^2\ .
\end{matrix} \label{eq:thirdProp}
\end{align}
Let  $\tilde \rho_{X^n W^t_0 W^t_1 \Pi B'''}$ be the shared quantum
state after the interactive hashing, where $B'''$ is Bob's part of
that state.
From (\ref{eq:thirdProp}) it follows that there exists a $C \in \{0,1\}$,
or more precisely, there exists an ideal state
$\tilde \sigma_{X^n W^t_0 W^t_1 \Pi B''' C}$ with
$\tilde \rho_{X^n W^t_0 W^t_1 \Pi B'''} = \tilde \sigma_{X
W^t_0 W^t_1 \Pi B'''}$, such that
\begin{align}
\begin{matrix}
\Pr_{\tilde \sigma} \left [ \hmin^{\varepsilon+4\delta}(\Pi(X^n)_{\Enc(W_{1-C}^t)}|W_0^t W_1^t \Pi B''')_{\tilde \sigma} \geq \left(\frac{\omega-1}{\omega}\right)\frac{\lambda n}{4} \right ] &\\
 \geq 1 - 32\delta^2\ . &
\end{matrix}\label{eq:goodsetinqv}
\end{align}
Note that Bob may use his quantum state during the interactive
hashing, but he cannot increase the probability of
~\eqref{eq:thirdProp} this way. Furthermore, any processing may only
increase his uncertainty. Let $\cA$ be the event that the inequality 
in the argument on the lhs. of~\eqref{eq:goodsetinqv} holds.
Let 
\[ \tilde \sigma_{X^n W^t_0 W^t_1 \Pi B''' C R_0 R_1} :=
 \tilde \sigma_{X^n W^t_0 W^t_1 \Pi B''' C} \otimes \tau_{\cR}
\otimes \tau_{\cR}
\] 
and let $S^\ell_0$ and $S^\ell_1$ be calculated as stated in the protocol. Using
the chain rule (see~\eqref{eq:chainrule}) and the fact that $(R_0,R_1)$ are independent, we get 
\begin{align*} 
\begin{split}\hmin^{\varepsilon+4\delta}(\Pi(X^n)_{\Enc(W_{1-C}^t)}| S_{C}^\ell C
R_0 R_1 W_0^t W_1^t \Pi B''', \cA)_{\tilde \sigma}\\
\geq \left(\frac{\omega-1}{\omega}\right)\frac{\lambda n}{4} - \ell - 1\;. 
\end{split}
\end{align*} 
Using privacy amplification (Theorem \ref{thm:PA}), we then have conditioned on the event $\cA$ that
\begin{align}
\begin{split}
\|\tilde{\sigma}_{S_{1-C},S_{C} C R_0 R_1 W_0^t W_1^t \Pi B'''} - 
\tau_{\sbin^\ell} \otimes \tilde{\sigma}_{S_{C} C R_0 R_1 W_0^t W_1^t \Pi B'''}\|_1
\\
\leq 2(\delta +
2\varepsilon+8\delta)\ , \hspace{5ex}&
\end{split}
\end{align}
since
\[\left(\frac{\omega-1}{\omega}\right)\frac{n \lambda}{4} - 2\ell - 1\geq 2 \log 1/\delta = 2 \cdot
\frac{ \lambda^2 m}{512 \omega^2}\;,\] which follows from
\[\ell \leq \left(\frac{\omega-1}{\omega}\right)\frac{\lambda n}{8}  -  \frac{ \lambda^2 m}{512 \omega^2} - \frac{1}{2}\;.\]
Let $B^* := (R_0 R_1 W_0^t W_1^t \Pi B''')$ be Bob's part in the output state. Since
$\Pr[\cA] \geq 1- 32 \delta^2$, we get
\[
 \tilde \sigma_{S_{1-C} S_C B^* C} \approx_{32 \delta^2 +9\delta +
2\varepsilon} \tau_{\{0,1\}^\ell} \otimes \tilde \sigma_{ S_C B^* C}
\]
and
\[
 \tilde \sigma_{S_0 S_1 B^*} = \tilde \rho_{S_0 S_1 B^*}\;.
\]
Since $\delta^2 \leq \delta$, this implies the security condition for
Alice, with a total error of at most
$41 \delta + 2\varepsilon$.
\end{proof}

Finally, we show that the protocol is correct when Alice and Bob are both honest.

\begin{lemma}[Correctness] 
\label{lem:FROT-correctness}
Protocol WSE-to-FROT satisfies correctness with an error of
\[ 43 \cdot 2^{-\frac{\lambda^2}{512 \omega^2\beta} n}\;.\]
\end{lemma}

\begin{proof}
Let $\xi := 2^{-n/16}$. We have to
show that the state $\tilde \rho_{S_0^\ell S_1^\ell C Y^\ell}$ at the
end of the protocol is close to the given ideal state $\tilde
\sigma_{S_0^\ell S_1^\ell C S^\ell_C}$.
Using the Hoeffding bound~\cite{Hoeffding}, the probability that a
random subset of $[n]$ has less than $n/4$ elements is at most
$\exp(-n/8) \leq \xi$.  Hence the
probability that Bob has to pad~$s$ with $0$s (which are likely to be incorrect)
when both parties are honest is is at most $\xi$. Let $\cA$ be the event that this does not happen.
It remains to show that the state $\tilde \rho_{S_0^\ell S_1^\ell C
Y^\ell \mid \cA}$ is close to the given ideal state
$\sigma_{S_0^\ell S_1^\ell C S^\ell_C}$. Note that the correctness
condition of WSE ensures
that the state created by WSE is equal to $\rho_{X^n \cI X_{\cI}} =
\sigma_{X^n \cI X_{\cI}}$, where $\sigma_{X^n \cI} = \tau_{\{0,1\}^n}
\otimes \tau_{2^n}$. Since $\cI_0$ and $\cI_1$ are chosen independently of
of $X^n$, $X_{\cI_0}$ and $X_{\cI_1}$ 
have a min-entropy of $n/4$ each. Since $\ell \leq n/8 \leq n/4 -
2 \log 1/\xi$, it follows from Theorem~\ref{thm:PA} that $S^\ell_C$ 
is independent and $\xi$-close to uniform. Since dishonest Bob is only more
powerful than honest Bob, we furthermore have from Lemma~\ref{lem:FROT-secAlice} that also $S^\ell_{1-C}$
is independent and uniform except with an error of at most 
$\hat{\eps} = 41 \cdot 2^{-\frac{\lambda^2}{512 \omega^2 \beta} n}$, where we used the fact that
Bob is also honest during weak string erasure ($\eps = 0$).
Finally, by the same arguments as in Lemma \ref{lem:FROT-secBob} we have that $C$ is uniform
and independent of $S^\ell_0$ and $S^\ell_1$. Hence,
\[\rho_{S_0^\ell S_1^\ell C \mid \cA} \approx_{\xi + \hat{\eps}} \sigma_{S_0^\ell
S_1^\ell C}\;.\]
Since the extra condition on the permutation $\Pi$ implies that Bob
can indeed calculate~$\Pi(X^n)_{\Enc(W)}$  from $X_{\cI}$, we have that
$Y^\ell = S_C^\ell$. Using $\Pr[\cA] \geq 1 - \xi$, we get
\[\rho_{S_0^\ell S_1^\ell C Y^\ell} \approx_{2 \xi + \hat{\eps}} \sigma_{S_0^\ell
S_1^\ell C S^\ell_C}\;.\]
Finally, $\lambda \leq 1$, $\beta > 1$ and $\omega \geq 2$ 
give us $1/16 > \lambda^2/(512 \omega^2\beta)$. Adding up all errors and noting that
\[2 \cdot 2^{-\frac{1}{16}n} \leq 2 \cdot 2^{-\frac{\lambda^2}{512 \omega^2\beta} n}\;,\]
gives our claim.
\end{proof}
\section{Conclusions and open problems}\label{sec:extensions}\label{sec:conclude}
We have shown that secure bit commitment and oblivious transfer can be obtained with unconditional security in the \nsw. 
We have connected the security of
our protocols to the information-carrying capacity of the noisy channel describing the malicious party's storage. We found a natural tradeoff between the (classical) capacity of the storage channel and the rates at which oblivious transfer and bit commitment 
can be performed: higher noise levels lead to stronger security.

The connection between capacities of channels and security turns out to be directly applicable to a number of settings of practical interest. At the same time, our work raises several immediate questions concerning the exact requirements for security in the \nsw. It has already led to follow-up work: our technique of relating security to a coding problem has been used to construct another, simpler, protocol for oblivious transfer~\cite{chris:newOT}, albeit
at the expense of requiring significantly more noise in memory to achieve security. Other channels have been shown to satisfy the strong converse property, and hence
lead to security in our model~\cite{hayashi:newChannels}. Alternate forms of weak string erasure using high dimensional states have been investigated using our techniques to
show that in the limit of large $n$, security in bounded-storage model holds as long as a constant fraction of transmitted states is lost (i.e., $\nu < 1$)~\cite{prabha:highDim},
and the security of an eavesdropper with a noisy memory device in QKD was investigated~\cite{bocquet:qkd}. A practical implementation is in progress~\cite{erven:qkd}.

\noindent

\textbf{Extending security:}
Clearly, it is desirable to extend the security guarantee to a wider range of noisy channels. 
The
limiting factor in obtaining security from a noisy storage described by $\cF = \cN^{\otimes \nu n}$ was the fact
that we require 
the sufficiency condition $C_{\cN}\cdot \nu < 1/2$ to hold (see Corollary~\ref{cor:mainstatement}),
where~$\nu$ is the storage rate and $C_{\cN}$ is the classical capacity of $\cN$. 
The constant~$1/2$ is a result of using BB84-states, and stems from a corresponding 
uncertainty relation using post-measurement information~\cite{maassen:entropy}.
It is a natural question whether we can go beyond this bound using BB84-encodings.

For channels with small classical capacity, our work reduces security to proving a strong converse for coding. 
Of considerable practical interest are continuous-variable channels: our results are also applicable in this case, 
given a suitable bound on the information-carrying capacity.  

A more challenging question is to extend security to entirely different classes of channels than considered here. Our results are currently restricted to  channels without memory. Possibly the most important class of channels to which our results do not apply are those with high classical capacity. 
This includes for example the dephasing channel whose classical capacity is~$1$. Security tradeoffs for such a channel are known~\cite{amir:personal} for the case of individual storage attacks~\cite{prl:noisy}. 
For the fully general case considered here, it is not a priori clear whether small classical capacity is a 
necessary condition for security: Our security proof overestimates the capabilities
of the malicious party by expressing his power
purely by his ability to preserve classical information. 
Completely different techniques may be required to address this question.

Another way to extend our security analysis is to combine our protocols with computationally secure protocols
to achieve security if the adversary either has noisy quantum storage \emph{or} is computationally bounded.
This can be achieved by using \emph{combiners} (see \cite{Herzbe05,HKNRR05,MePrWu07}).
For oblivious transfer, the same can be achieved using the techniques of \cite{BBCS92,Yao95,CDMS04,DFLSS09},
which only requires the use of a computationally secure bit commitment scheme.

\smallskip 
\noindent
\textbf{Limits for security:}
We have found {\em sufficient } conditions for security in the \nsw. For concrete channels, these conditions give
regions in the plane parametrized by the storage rate and the noise level (cf.~Figure~\ref{fig:depolarizingchannelanalysis}) where security is achievable. Establishing outer bounds on the achievability region is an interesting open problem.  Corresponding {\em necessary} conditions could become practically relevant as technology advances. 

Note that when the adversarial player is restricted to individual storage attacks, the optimal attacks are known~\cite{noisy:robust}. It is an open problem whether the fully general coherent attacks  considered here  actually reduce the achievability region. In contrast, both kinds of attacks 
are known to be equivalent in QKD~\cite{renato:diss}.

\smallskip 
\noindent
Our work is merely a proof of principle. For practical realizations of our protocols, the following issues need to be addressed:

\smallskip 
\noindent
\textbf{Efficiency:}
One can reduce the 
amount of classical computation and communication needed to execute our protocols by using techniques from derandomization. 
In particular, we could use the constant-round interactive hashing protocol and the efficient encoding of 
subsets from~\cite{ding2}, randomness-efficient samplers (see e.g.,~\cite{goldreich:samplers}), and
extractors (see e.g.~\cite{Shaltiel,rb:oneBit,tashma}) instead of two-universal hash functions.

In practice, both the security parameter~$\eps$ and the number~$\ell$ of bits in the commitment or oblivious transfer are fixed constants. Savings in communication  may then be obtained by using alternative uncertainty relations (i.e., generalizations of~\eqref{eq:postmeasurementminentropyuncertainty}, which is tight~\cite{ww:pistar} for $\varepsilon=0$).
\smallskip 

\noindent
\textbf{Composability:}
We have shown security of oblivious transfer and bit commitment with respect to security definitions that are
motivated by composability considerations: This should ensure that the protocols remain secure 
even when executed many times e.g., sequentially. It is, however, an open problem to
show formal composability in our model as has been done in the setting of bounded-storage~\cite{ww:compose,fs:compose}. To this end, 
a composability framework for our setting needs to be established.

\smallskip 
\noindent
\textbf{Robustness:}
We have considered an idealized setting where the operations of the honest parties are error-free. In particular, the communication channel connecting Alice and Bob was assumed to be noiseless. 
In real applications, both the BB84-state preparation by (honest) Alice, the communication, and the measurement of (honest) Bob will be 
affected by noise. To guarantee security even in such a setting, we can apply the error-correction techniques of~\cite{noisy:robust}. 
However, it remains to determine
the exact tradeoff between the amount of tolerable noise of the communication channel (parametrized e.g., by the bit error rate) and the amount of noise in the malicious player's storage device~\cite{marcos:practical}.

We conclude with a few speculative remarks on potential applications of our work. 
Note that, in contrast to key distribution, general two-party computation is 
also interesting at short (physical) distances. 
An example is the problem of secure identification~\cite{bounded:secureId}, where Alice wants to identify herself to Bob
(possibly an ATM machine) without ever giving her password away. 
Our approach could be extended to realize this primitive in a similar way as in~\cite{noisy:robust}. It would
be interesting to find a new and more efficient protocol based directly on weak string erasure. 
The setting of secure identification is especially suitable for our model, 
since the short distance between Alice and Bob implies that their communication channel is essentially error-free.
At such short range, we could also use visible light for which much better detectors exists than are presently used in 
quantum key distribution. Note that Alice only needs to carry a device capable of generating BB84-states and allowing her to
enter her password on a keypad. This device does not need to store any information about Alice herself and 
hence each user could carry an identical device which is 
completely exchangeable among different (trusted) users at any time. In particular, this means
means that Alice's password is not compromised even if the device is lost.
Finally, note that 
Alice's technological requirements are minimal: She only needs a device capable of generating BB84-states. This
could potentially be small enough to be carried on a key chain. 

\section*{Acknowledgments}
We thank Marcos Curty, Andrew Doherty, Amir Kalev, Hoi-Kwong Lo, Oded Regev, John Preskill 
and Barbara Terhal for interesting discussions. We also thank Christian Schaffner for discussions and comments on an earlier draft, and
Dominique Unruh for pointing out a flaw in the proof of
Theorem 3.5 in an earlier version of the paper, as well as for various other
useful suggestions.
RK and SW are supported by NFS grants PHY-04056720 and PHY-0803371. SW was also supported by the National Research Foundation and the Ministry of Education, Singapore. JW is supported by the U.K.\ EPSRC, grant EP/E04297X/1 and the Canada-France NSERC-ANR project FREQUENCY.
Part of the work done while JW was at University of Bristol (UK) and visiting Caltech (Pasadena, USA).

\appendix

\section{Proofs for min-entropy sampling}

\subsection{The parameters for sampling -- proof of Lemma~\ref{lem:samplingmodified}\label{sec:sampler}}
For the proof of Lemma~\ref{lem:samplingmodified}, we first recall the definition of a {\em sampler}: 
\begin{definition}
An $(m,\xi,\gamma$)-averaging sampler is a probability
distribution over subsets $\cS\subset [m]$ with the property that for
all
$(\mu_1,\ldots,\mu_m)\in [0,1]^{m}$ we have
\begin{align*}
\Pr_{\cS} \left[\frac{1}{|\cS|}\sum_{i\in\cS}\mu_i \leq
\frac{1}{m}\sum_{i=1}^m\mu_i -\xi \right]\leq \gamma\ .
\end{align*}
\end{definition}
Choosing subsets of a fixed size at random is a prime example of a sampler; this is the sampler we will use. The parameters of this sampler 
are as follows:
\begin{lemma}\label{lem:subsetsampler}
Let $s<m$ and let $P_{\cS}$ be the uniform distribution over subsets
$\cS\subset [m]$ of size $|\cS|=s$. Then $P_{S}$ is an $(m,\xi,2^{-s\xi^2/2})$-sampler
for every $s > 0$ and $\xi \in [0,1]$.
\end{lemma}
\begin{proof}
Fix $s>0$ and $\xi\in [0,1]$. In~\cite[Lemma~2.2]{kr:sampling},~$P_\cS$ was shown to be a $(m,\xi,e^{-s\xi^2/2})$-sampler. The claim then follows from the fact that~$e^{-s\xi^2/2}\leq 2^{-s\xi^2/2}$.
\end{proof}
 
Replacing $h_{\min}$ by $\hmin$,
 the following lemma follows directly from~\cite[Lemma
6.15 and Lemma~6.20]{kr:sampling}. The proof follows the same step as the proof of Theorem~6.18 in~\cite{kr:sampling}. 

\begin{lemma}\label{lem:minentropysampling}
Let $\rho_{Z^mQ}$ be a cq-state, where $Z^m=(Z_1,\ldots,Z_m)$ with $Z_i\in \sbin^\beta$, and let
$P_{\cS}$ be an $(m,\xi,\gamma)$-averaging sampler supported on subsets~$\cS$ of size $s=|\cS|$. Then
\begin{align*}
\Pr \left [ \frac{\hmin^{4 \gamma^{1/4}}(Z_\cS|\cS Q)}{s\beta} \geq
\frac{\hmin(Z^m|Q)}{m\beta}   - c \right ] \geq 1 - \sqrt{\gamma}
\end{align*}
 where
\begin{align*}
c =  \frac{ \log 1/\gamma}{2m\beta}  + \xi + 2 \kappa \log 1 / \kappa
\end{align*}
and
 $\kappa = \frac{m}{s \beta} \leq 0.15$.
\end{lemma}
Specializing Lemma~\ref{lem:minentropysampling} to  the sampler defined by Lemma~\ref{lem:subsetsampler} gives the following statement.

\begin{lemma}[Min-entropy block sampling~\cite{kr:sampling}]\label{lem:blocksamplingmodified}
Let $\rho_{Z^mQ}$ be a cq-state as in Lemma~\ref{lem:subsetsampler}, and let
\begin{align*}
\frac{\hmin^{\varepsilon}(Z^m|Q)}{m\beta} \geq \lambda 
\end{align*}
be a lower bound on the smooth min-entropy rate of~$Z^m$ given~$Q$.
Let $\omega \geq 2$ be a constant, and assume $s,\beta\in\mathbb{N}$ are such that 
\begin{align}
s\geq m/4\qquad\textrm{and }\qquad \beta\geq \max \left
\{67,\frac{256 \omega^2}{\lambda^2} \right \}\ .\label{eq:sbetacondsec}
\end{align}
Let $P_\cS$ be the uniform distributions over subsets of~$[m]$ of
size~$s$.  Then
\begin{align*}
\Pr_{\cS}\left[\frac{\hmin^{\varepsilon+4\delta}(Z_{\cS}|Q)}{s\beta}\geq
\left(\frac{\omega-1}{\omega}\right)\lambda \right]&\geq 1-\delta^2
\end{align*}
where $\delta=2^{-m \lambda^2/(512 \omega^2)}$.
\end{lemma}
\begin{proof}
Because of the definition of smooth min-entropy and the fact that partial traces do not increase distance, it suffices to establish the claim for~$\eps=0$.
By Lemma~\ref{lem:subsetsampler} and
Lemma~\ref{lem:minentropysampling},  we have
\begin{align}
\begin{matrix}
\Pr \left [ \frac{\hmin^{ 4\cdot 2^{-s\xi^2/8}}(Z_\cS|\cS Q)}{s\beta} \geq
\frac{\hmin(Z^m|Q)}{m\beta}   - c \right] \geq 1 - 2^{-s\xi^2/4}\\
\textrm{where}\qquad  c =  \frac{s \xi^2}{4m\beta}  + \xi + 2 \sqrt{\kappa}
\end{matrix}\label{eq:firstpartproof}
\end{align} 
if  $\kappa = \frac{m}{s \beta} \leq 0.06$. Here we used the inequalities
\begin{align*}
\kappa\log 1/\kappa&\leq \sqrt{\kappa}\qquad\textrm{for }\kappa\leq 0.06\ ,\\
\beta & \geq 1\ , \qquad s \leq m\qquad\textrm{ and }\qquad \xi\leq 1\ .
\end{align*}
Note that the condition $\kappa\leq 0.06$  is satisfied if
 $s\geq m/4$ and $\beta\geq 67$.
Setting $\xi :=\textfrac{\lambda}{(4\omega)}$ and using $s\geq m/4$ again, we get
for
$256 \omega^2/\lambda^2\leq \beta$ that
\begin{align*}
c&\leq \frac{\lambda^2}{64 \omega^2 \beta} + \frac{\lambda}{4\omega} + 4\sqrt{\frac{1}{\beta}}\\
&\leq \frac{\lambda^4}{2^{14} \omega^4} + \frac{\lambda}{4 \omega} + \frac{\lambda}{4\omega}\\
&\leq \frac{\lambda}{2 \omega} + \frac{\lambda}{2 \omega} = \frac{\lambda}{\omega} 
\end{align*}
In particular, this implies that
\begin{align}
\frac{\hmin(Z^m|Q)}{m\beta}-c&\geq \lambda - c \geq \left(\frac{\omega-1}{\omega}\right)\lambda\ .\label{eq:lastpartproof}
\end{align}
Combining~\eqref{eq:firstpartproof},~\eqref{eq:lastpartproof} 
with $s\xi^2=s\lambda^2/(16 \omega^2)\geq m\lambda^2/(64 \omega^2)$ and $\delta=2^{-m\lambda^2/(512 \omega^2)}$ gives the claim.
\end{proof}

Instead of grouping the $m \beta$ bits into $m$ blocks $Z_i \in \{0,1\}^\beta$, let us now look at a normal bit string $X^{\beta m} \in \{0,1\}^{\beta m}$. The following lemma has been proven in \cite{bitwiseSampling} and shows that the bound in Lemma~\ref{lem:blocksamplingmodified} can also be achieved if the sample is a subset of size $s \beta$ chosen bitwise uniformly at random.

\begin{lemma}[Min-entropy block sampling implies bit sampling \cite{bitwiseSampling}]\label{lem:bitwise}
The bound of Lemma~\ref{lem:minentropysampling} also applies if the sample is chosen bitwise uniformly. More generally, assume that $s,m,\beta,\lambda,\lambda',\varepsilon,\varepsilon',\delta'$ are such that the following holds:
For all cq-states $\rho_{Z^mQ}$ with $Z^m=(Z_1,\ldots,Z_m)$, $Z_i\in\{0,1\}^\beta$ and $\frac{\hmin^\varepsilon(Z^m|Q)}{m\beta}\geq \lambda$,
we have 
\begin{align}
\Pr_{\cT}\left[\frac{\hmin^{\varepsilon'}(Z_\cT|Q)}{s\beta}\geq \lambda'\right]\geq 1-\delta'\ ,\label{eq:lowerboundblocksamplerT}
\end{align}
where $P_{\cT}$ is the uniform distribution over subsets of $[m]$ of size~$s$.  Let $P_{\cS}$ be the uniform distribution over subsets of $[m\beta]$ of size~$s\beta$. We then have
\begin{align*}
\Pr_{\cS}\left[\frac{\hmin^{\varepsilon'}(X_\cS|Q)}{s\beta}\geq \lambda'\right]\geq 1-\delta'\ 
\end{align*}
for all cq-states $\rho_{X^{m\beta}Q}$ with $X^{m\beta}\in\sbin^{m \beta}$ and  $\hmin^\varepsilon(X^{m\beta}|Q)\geq \lambda$. 
\end{lemma} 
 
\begin{proof}
Let $\rho_{X^{m\beta } Q}$ be a cq-state where $X^{m\beta} \in \{0,1\}^{m \beta}$. Let $\cS \subset [m \beta]$ be chosen uniformly at random from all subsets of size $s \beta$ and let $\cT \subset [m \beta]$ be a random subset of size $s \beta$ chosen blockwise as in~\eqref{eq:lowerboundblocksamplerT} (that is, after rearranging~$X^{m\beta}=Z^m=(Z_1,\ldots,Z_m)$ into a collection of $\beta$-bitstrings). Let $\Pi$ be a permutation chosen uniformly at random, but such that it maps all elements in~$\cS$ into~$\cT$.
Strong subadditivity (Theorem~3.2.12 in \cite{renato:diss}) implies
\begin{align*}
 \hmin^{\eps'}(X_{\cS} | \cS Q )
 &\geq \hmin^{\eps'}(X_{\cS} | \cS \Pi Q ) \\
 &= \hmin^{\eps'}(\Pi(X^{m\beta})_{\cT} | \cT \Pi Q ) \;.
\end{align*}
Note that from $(\cS,\Pi)$ it is possible to calculate $(\cT,\Pi)$, and vice-versa.
Furthermore, since $\Pi$ is chosen independently of $\rho_{X^{m\beta}Q}$, we have
 \[\hmin^\eps(\Pi(X^{m\beta}) | \Pi Q) = \hmin^\eps(X^{m\beta} | \Pi Q) = \hmin^\eps(X^{m\beta} | Q)\;.\]
Since $\cS$ was chosen uniformly and independently of $\cT$ and~$\rho_{X^{m\beta}Q}$, $\Pi$
is independent of $\cT$ and $\rho_{X^{m\beta}Q}$.
Setting $Q' := (Q,\Pi)$, we can apply~\eqref{eq:lowerboundblocksamplerT}  to the state~$\rho_{\Pi(X^{m\beta}) Q'}$ and get a bound on $\hmin^{\eps'}(\Pi(X^{m\beta})_{\cT} | \cT \Pi Q )$, which then directly implies the same bound for $\hmin^{\eps'}(X_{S} | \cS Q )$.
\end{proof}

 Lemma~\ref{lem:samplingmodified} now immediately follows by combining Lemma~\ref{lem:blocksamplingmodified} with Lemma~\ref{lem:bitwise}.

\end{document}